\documentclass{article}

\usepackage{arxiv}
\usepackage[utf8]{inputenc} 
\usepackage[T1]{fontenc}    
\usepackage{hyperref}       
\usepackage{url}            
\usepackage{booktabs}       
\usepackage{amsfonts}       
\usepackage{nicefrac}       
\usepackage{microtype}      
\usepackage{lipsum}
\usepackage{graphicx}

\usepackage{natbib}
\usepackage{float}

\usepackage{amsmath,amssymb}
\usepackage{lmodern}
\usepackage{iftex}
\usepackage{enumerate}
\usepackage[utf8]{inputenc}
\usepackage{tabularray}
\usepackage{subcaption}
\usepackage{longtable}

\usepackage{colortbl}

\usepackage{xcolor}
\usepackage{longtable, book tabs, array}
\usepackage{calc} 

\definecolor{cb_blue}{RGB}{0,114,178}
\definecolor{cb_orange}{RGB}{230,159,0}
\definecolor{cb_green}{RGB}{0,158,115}
\definecolor{cb_pink}{RGB}{213,94,0}
\definecolor{cb_brown}{RGB}{128,133,133}
\definecolor{cb_purple}{RGB}{86,180,233}
\definecolor{cb_yellow}{RGB}{240,228,66}
\definecolor{cb_red}{RGB}{204,121,167}

\usepackage{etoolbox}
\usepackage{comment}
\usepackage{multirow}

\begin{document}

\author{
 Wenyi Lu \\
  Electrical Engineering and Computer Science\\
  University of Missouri\\
  Columbia, MO \\
   \And
 Enock Kasaadah \\
  Electrical Engineering and Computer Science\\
  University of Missouri\\
  Columbia, MO \\
\And
 S M Rakib Ul Karim \\
  Electrical Engineering and Computer Science\\
  University of Missouri\\
  Columbia, MO \\
\And
 Matt Germonprez \\
  College of Information Science and Technology\\
  University of Nebraska - Omaha\\
  Omaha, NE \\
\And  
 Sean Goggins \\
  Electrical Engineering and Computer Science\\
  University of Missouri\\
  Columbia, MO \\
}


\title{Open Source Software Lifecycle Classification: Developing Wrangling Techniques for Complex Sociotechnical Systems}

\maketitle


 
\begin{abstract}
Open source software is a rapidly evolving center for distributed work, and understanding the characteristics of this work across its different contexts is vital for informing policy, economics, and the design of enabling software. The steep increase in open source projects and corporate participation have transformed a peripheral, cottage industry component of the global technology ecosystem into a large, infinitely complex "technology parts supplier" wired into every corner of contemporary life. The lack of theory and tools for breaking this complexity down into identifiable project types or strategies for understanding them more systematically is incommensurate with current industry, society, and developer needs. This paper reviews previous attempts to classify open source software and other organizational ecosystems, using open source scientific software ecosystems in contrast with those found in corporatized open source software. It then examines the divergent and sometimes conflicting purposes that may exist for classifying open source projects and how these competing interests impede our progress in developing a comprehensive understanding of how open source software projects and companies operate. Finally, we will present an empirical, mixed-methods study demonstrating how to classify open-source projects by their lifecycle position. This is the first step forward, advancing our scientific and practical knowledge of open source software through the lens of dynamic and evolving open source genres. It concludes with examples and a proposed path forward. 
\end{abstract}

\section{Introduction}
\begin{quote}
    "We must consistently explore what is left dark by our current classifications and design classification systems that do not foreclose on rearrangements suggested by new social and natural knowledge forms. There are many barriers to this exploration. Not least among them is the barrier of boredom. Delving into someone else's infrastructure has about the entertainment value of reading the phone book's yellow pages." - ~\citet[p.321-322]{bowker_sorting_1999}
\end{quote}

Open Source Software (OSS) is emerging as a new way of working \citep{marlow_activity_2013, marlow_effects_2015, marlow_impression_2013, dabbish_social_2012}. It shows potential as a model for sustainable learning  \citep{fiesler_growing_2017, downing_socialization_2005, lin_developer_2017, howison_collaboration_2014}, decoupling work from a "workplace," and shaping how software changes the world around us \citep{bird_who?_2012, mens_ecosystemic_2016}. OSS also fosters a sense of agency over technology by lifting the lid off of software people are historically accustomed to having hidden \citep{agre_reinventing_1997, agre_changing_2001}. OSS projects are central components for the secure transmission of network packets to and from your computer, programming languages used to build web pages, and the temperature control on your smart thermostat. 

Ubiquity of use in contemporary society and transparency to inspection frame the impact of OSS. From ubiquity, it follows that when flaws are implemented in an OSS project, the potential for billions of people to be negatively affected ~\citep{bhartiya_linux_2016} via an OSS-carrying electronic device, banking transaction, or job task also grows. Understanding how and to what extent different classifications of OSS influence our day-to-day lives, how similar projects used by various firms might be identified, and the distinctive quality, sustainability, and health characteristics of individual open source projects embedded in distinct parts of our lives is essential. The dramatic growth of open source over the past decade makes understanding how to differentiate projects from each other a critical competency for individual developers, organizations, and society \citep{germonprez_eight_2018}. Techniques for differentiating OSS projects from each other on a large scale demand a set of wayfinding markers operationalized as classifications that are presently difficult even for experienced OSS contributors and impossible for the rest of us \citep{lumbard_are_2020}. 

Absent an established classification method for OSS, people rely heavily on their experiences, prior work, and current perspective to make sense of what they observe \citep{lumbard_are_2020, mcdonald_modeling_2014}. Challenges arise because people not already engaged in OSS view the broad umbrella of OSS as a mono-class, while those actively involved in OSS will arrive at a set of categories framed through years of work. The growth of the Linux Foundation over the past decade, for example, highlights the fact that a good deal of OSS now has significant corporate support in the form of engineering time and engagement.  

Viewing open source as a growing trend of corporate-communal engagement is one lens recently used to help explain the rapid growth of open source and speculate about ways for classifying projects with increased corporate engagement \citep{germonprez_theory_2017, germonprez_collectivism_2014}. For example, open source software foundations are emerging as a new type of organizational structure for coordinating work on open source that occurs across many workplaces, locations, and contexts ~\citep{benkler_wealth_2006, benkler_philosophy_2004, germonprez_theory_2017, germonprez_collectivism_2014, mcdonald_modeling_2014, germonprez_eight_2018}. Foundations, like the Linux Foundation (LF) ~\citep{linux_foundation_linux_2008}, for example, oversee strategically critical software infrastructures their members deem to be "technologies that are not market differentiating" ~\citep{smith_product_1956}. The LF started as a foundation to oversee the Linux Kernel but grew from one project to over 1,000 during the past decade, and the collective value of the code in Linux Foundation projects is estimated at roughly \$16 billion\footnote{\url{https://www.linuxfoundation.org/about/}}. Today, the LF is an umbrella foundation for several others, focused on particular segments of the software market, including the Cloud Native Computing Foundation (CNCF) and more than 50 infrastructure-focused initiatives like Auto Grade Linux and Zephyr-RTOS (A real-time operating system) for the "Internet of Things."

Research to date is insufficient and, to some extent, fractured across more purely technology-oriented fields to address the challenges associated with helping society, industry, organizations, groups, or individuals navigate the OSS Ocean. Most prior research on open source examines small subsets of open source projects, typically using convenience sampling that originates with social connection, technology awareness, or proximity to firms actively engaged in corporate-communal partnerships. 

The problems with \textbf{convenience sampling} are twofold: (1) it reinforces our cognitive biases \citep{harding_cognitive_2004, azizi_is_2018} which, in turn, (2) limits the potential for insights and advances that push science forward \citep{kuhn_structure_2012}. Alternately, the OSS Ocean contains tens of thousands of active projects and millions of public repositories, each with significant contributors. Whatever set of OSS projects we examine is unlikely to be representative of the different kinds of people, projects, and organizational material that make up open source today. These types of "convenience samples" do not tell a complete story, and perhaps more significantly, we cannot accurately characterize which part of the larger OSS story they tell\citep{germonprez_eight_2018} without some reasonable inventory of classification schemes across the whole ocean of OSS ecosystems \citep{bowker_sorting_1999}. To match our understanding with the diversity of ecosystems in the OSS Ocean \citep{daniel_effects_2013}, we must identify projects with similarities along some measurable dimensions. 

It is critical for our understanding of open source and its implications for daily life work, and as an exemplar of new ways to learn that we develop a framework for identifying project similarities and differences; we need a clear means for operationalizing social comparisons between projects \citep{festinger_theory_1954} and, in turn, advance our knowledge of the effects of technology on life itself \citep{lumbard2024empirical}. OSS projects rely on a flood of fresh ideas, which enable rapid adaptation to a rapidly changing technology landscape. However, as with any social situation, there can be challenges between established group members focused on current goals and newcomers exploring projects. These dynamics are political \citep{germonprez_eight_2018} regarding competition for a limited time and other resources. Political situations, in turn, are fundamentally social relations problems where skilled communicators tend to excel \citep{agre_reinventing_1997}. 

Despite open source's social and political realities, most previous studies have looked closely at other characteristics that could be used for classification. These efforts include predictive models aimed at estimating project lifespan \citep{germonprez_collectivism_2014}, community involvement \citep{benkler2004commons}, and sociotechnical elements \citep{daniel_effects_2013}. Prior work also includes efforts to classify OSS along dimensions of governance style and license type \citep{plakidas_how_2016, singh_network_2011, stewart_impact_2006, steinmacher_systematic_2015, di_bella_multivariate_2013, goggins_group_2013, goggins_creating_2013, mcdonald_modeling_2014}. Few prior efforts have attempted to look across open source at scale to tease out how social factors can be operationalized by analyzing the substantial digital trail OSS leaves in its wake. Open source is not a closed box nor a monolith — it is a dynamic, context-dependent collection of interconnected ecosystems.  

The present study outlines a first step for understanding different OSS project dynamics by classifying them according to their position in a software lifecycle, using an established set of already classified projects. Our findings reveal that the most critical factor influencing project classification is an array of sociotechnical considerations, including the number of new contributors, star count, average commits per pull request, and software dependency count. Definitions for measuring these sociotechnical factors are pulled from the CHAOSS OSS project\footnote{\url{https://chaoss.community}}, and the classification accuracy in our findings is 90.3\%.

The remainder of this paper is structured as follows. Section two describes prior work on classification and wayfinding in OSS. Section three details the research questions and the methodology used to collect and analyze data. Section four presents the findings. Section five discusses the implications of these findings for OSS project sustainability. Finally, Section Six concludes the paper and suggests directions for future research. 

\section{Prior Work: Classification and Wayfinding in Open Source Research}
The problem of classifying specific "things" is widely understood through examples from medicine, the natural world, literature, and many other domains \citep{bowker2000sorting, devitt_generalizing_1993}. Finding robust, functional classification dimensions in any domain is difficult for myriad reasons, including challenges of large-scale infrastructure, the complexities of biographical perspective, and the diversity of work practice \citep{bowker_sorting_1999}. These general classification challenges observed by Bowker and Star \citep{bowker_sorting_1999} decades ago have clear corollaries in open source: OSS projects and the ecosystems they live in are embedded in a complex array of interconnected software and technology infrastructures, they have dozens of larger than life personalities and work practices are as diverse as the sea of humanity who define its character. 

Consider, for a moment, the offhand pitfalls you might see if we organized OSS projects around programming language. Are all Python projects similar? Or, perhaps, technology domains like file systems, web browsers, and web frameworks. Some projects will appear authentically identical, and others will simply not fit. How does one manage complexities like these when they are not fully described and generally understood by people engaged in open source work?
Classification forces the negotiation of shared human understanding and expresses it as piles of similar things. The essential components of the human body are classified by medicine and the high-level subject matter of human expression by the Dewey Decimal System. Automated classification systems identify and classify e e-mail messages as spam or not spam~\citep{meyer_spambayes:_2004}. Open source software is a more dynamic and continuously evolving phenomenon, making the likelihood of a similarly coherent classification system seem unlikely, and negotiating with those examples as a target is improbable. Instead of classification as a goal, it may be helpful to consider open source software types as situated in culture, project practices, situation, and their ecosystem of interdependent projects and products \citep{devitt_writing_2004}.

\subsection{Open Source Project Classification}
Open source projects are often classified according to the type of software they develop~\citep{borgespopularity2016, cosentino_systematic_2017}, commercial or non-commercial activity~\citep{cosentino_systematic_2017, kalliamvakou2015open}, and through badging that may signal quality~\citep{trockman_adding_2018}. Further, much work on classification has been on automated classification systems that have been used to classify repositories by application and programming language and have even identified recurrent patterns of developers across many different open source projects~\citep{di_bella_multivariate_2013,ugurel_whats_2002,ma_automatic_2018, kawaguchi_mudablue:_2006}. Others have worked to identify factors related to the sustainability of open source projects, exploring the different knowledge-sharing techniques and mechanisms~\citep{jafari_navimipour_knowledge_2016,cosentino_systematic_2017} that inhabit sustainable projects. Further, projects viewed as complex are less sustainable~\citep{grigorio_using_2014}.

Organization, business, and software structures can be conceptual frameworks for classifying open source projects~\citep{manikas_revisiting_2016}. Previous studies show that organizational structures and project architectures correlate with project code complexity~\citep{murgante_study_2014}, suggesting that sociotechnical processes sometimes shape technical characteristics. Further, organization structures affect merge activity with projects classified as "non-centralized," corresponding to high merge rates ~\citep{mcdonald_modeling_2014}. Other studies go on to characterize how decentralization helps to explain technical and non-technical barriers to contribution ~\citep{steinmacher_systematic_2015} and why contributions to projects are sometimes not merged~\citep{steinmacher_almost_2018}. Classification of these social activities reveals patterns of activity that predict contributor participation~\citep{gharehyazie_developer_2015}. Projects are also sometimes classified by the type of software they create \citep{borgespopularity2016}, their commercial or non-commercial activity \citep{cosentino_systematic_2017}, and through quality signals such as badging \citep{trockman_adding_2018}. Finally, researchers have created models predicting the lifetime of OSS projects using analysis of these elements and their relationships. These models help project managers make wise decisions to preserve project health and advance long-term sustainability \citep{lumbard2024empirical}.

\subsection{Machine Learning and Community Characteristics for Classification}
Leveraging machine learning and natural language processing (NLP), automated classification systems group repositories by application, programming language, and developer patterns \citep{di_bella_multivariate_2013, ugurel_whats_2002, ma_automatic_2018}. Improving OSS project durability depends on knowing technical and non-technical obstacles to contributions. Studies on the reasons behind the occasionally declining acceptance of contributions have found trends that forecast contributor involvement \citep{steinmacher_systematic_2015, steinmacher_almost_2018}. These realizations help to create a friendly atmosphere for new members, which is vital for the long-term survival of OSS projects.

OSS projects' longevity depends critically on knowledge-sharing systems and approaches. Effective knowledge-sharing mechanisms and their favorable influence on project sustainability have been underlined by systematic reviews \citep{jafari_navimipour_knowledge_2016,cosentino_systematic_2017}. Furthermore, research on the complexity of OSS projects has shown that their sustainability may suffer, emphasizing the importance of reasonable project architecture \citep{franco2017open}.

OSS research has been advanced with an eye toward the predictive modeling of project sustainability. For instance, Liao et al. \citep{liao2020productivity} presented a productivity model for examining elements affecting the state of software ecosystems on GitHub. Predictive models, including the HSPM model, have given a fresh understanding of OSS project continuation and health hazards, thus supporting better project management and strategic planning \citep{liao2023hspm}. Systematic literature evaluations underline the constant interaction and management requirement to preserve ecosystem health by including software ecosystems in more general corporate and digital settings \citep{burstrom2022ecosystems}. Estimating the lifetime and performance of OSS projects have shown success for predictive models, including programming language, project size, and community involvement \citep{liao2019prediction}.

\subsection{Project Lifecycle Classification}
It is essential to know what stage a project is at in the lifecycle chain to decide whether it is good to use or not \citep{liao2023hspm}. There has been increasing research about open source software life cycles and how they are affected by Software Ecosystems \citep{liao2020productivity}. In their comprehensive literature review, Burström et al. \citep{burstrom2022ecosystems} concluded that ecosystems are too complex to have one definitive description. This, in turn, has led to unclear divisions of life cycle phases \citep{liao2019prediction} of projects and unified definitions for characteristics that make up the projects in those ecosystems. There is a need for extensive study on factors that may help define project characteristics extensively to obtain the correct data for training classification models that speak to the lifecycle and sustainability of projects \citep{burstrom2022ecosystems}. 
    
Several attempts have been made to develop generic project life cycles and classification models to help differentiate projects; some have shortcomings and strengths in how they were implemented and benchmarked. Liao et al. \cite{liao2019prediction} attributed software life cycles to internal and external characteristics, which included choice of programming language, number of files, label format of the project, and relevant membership expressions using the source ecosystem projects. However, they use the label format as a classification feature. Yet, it does not have a standard in the ecosystem, i.e., anyone can come up with whatever label they see fit for their project depending on the labeling purpose, making it a weak general classification feature. They also attribute the 'core developers' feature to how many followers a developer has rather than how many contributions they have made to a specific project. This narrow description and criteria potentially exclude significant contributors whose core contribution is not based on how many followers they have. Also, given the amount of data on GitHub, we realized that more features would help classify projects and influence a better classification model than using only core developers. 

Kuwata and Miura \cite{kuwata2015study} predict the stage of projects in the Apache Software Foundation and OpenStack Foundation ecosystems using only project commits count, which may be a potentially misleading metric when used alone to predict the stage of a project. The number of commits at a particular time may not directly indicate the project lifecycle stage, but we believe it is a potential influence.    Liao et al. \cite{liao2020productivity} focus on how productivity determines the stage of projects in the GitHub ecosystem; even though they define productivity as the ability of the software ecosystem to produce ecological products, they narrow the factors affecting productivity only to Pull Requests, yet other factors would influence productivity. Their study also has a narrow time frame of only one month, which is very narrow for measuring the productivity of projects older than one year. In our study, we focus on using more than one metric to predict the stage of the lifecycle of projects and also look at how those different metrics come together to influence the classification model.

\subsection{Creating Classifiers Using Known Project Lifecycle Stages}
No prior work leverages a widely used and lifecycle-classified collection of projects like those in the CNCF\footnote{https://www.cncf.io}. This paper focuses on known lifecycle states for CNCF projects, intending to use the models developed to classify other projects into a distinct lifecycle state. CNCF explores, incubates, and maintains OSS cloud-native infrastructure and projects. These projects\footnote{https://www.cncf.io/projects/} are divided into three classifications, depending on how they fit the descriptions below:
\begin{enumerate}
\item \textit{Sandbox projects}: These projects have qualified to be part of CNCF and have not yet gained significant traction. Some of these projects include ContainerSSH, Athens, Antrea, and Curiefense. 
\item \textit{Incubating projects}: These projects are used in production by a small number of users, have a healthy pool of contributors but not as many as those in the graduated projects, and are slightly unstable. Some of these projects include gRPC, Backstage, Keycloak, and Knative. 
\item \textit{Graduated projects}: These are stable, widely adopted, production-ready projects and attract thousands of contributors. Some of these projects include Kubernetes, Linkerd, Prometheus, and Harbor. 
\end{enumerate}

\section{Research Questions, Data, and Methods}

\subsection{Research Questions}

In this paper, we will answer the following research questions:
\begin{enumerate}
    \item What aspects of open-source software project activity, including community engagement, responsiveness, and project complexity, serve as the most significant indicators of a project’s lifecycle stage?
    \item How do sociotechnical factors shape the classification boundaries, progression patterns, and interpretability of models predicting OSS project lifecycle stages?
\end{enumerate}

\subsection{Data}
We leveraged a comprehensive set of features derived from project data on  GitHub\footnote{https://github.com}, focusing on Cloud Native Computing Foundation (CNCF)\footnote{https://www.cncf.io/} projects. Table \ref{Tab:InitialRun} describes some of the core metric data gathered by Augur \citep{goggins_augur:_2021}.

\begin{table}[h]
\centering
\begin{tblr}{
  colspec={|Q||Q|},
  rows={halign=r},
  column{1}={halign=r},
  column{2}={halign=l},
  row{1}={halign=c},
  row{2}={halign=l},
}
 \hline
 \SetCell[c=2]{} Description of Data Scale & \\
 \hline
   Number of Rows   & OSS Production Construct\\
 \hline
106,233 & repositories  \\
933,312,263 & commit files changed \\
13,987,099 & pull requests \\
195,227,162 & pull request files \\
54,210,637 & pull request commits \\
29,680,212 & pull request comments \\
27,111,795 & pull request reviews \\
20,240,801 & pull request review comments \\
7,407,042 & issues \\
23,944,706 & issue comments \\
 \hline
\end{tblr}
\caption[Description of Data]{\label{Tab:InitialRun}Description of Data gathered by Augur \citep{goggins_augur:_2021} and used for analysis and modeling. Data includes all project activities for each project, beginning with the date its Git repository started through May of 2024. Projects initiated before the establishment of Git platforms like GitHub typically contain the commit records for their full history and all other data from the date they joined a platform. The first commit date in this data set is January 1, 2001, four years before the creation of Git and six years before the dawn of GitHub. These commits are migrated from earlier software version control systems.}
\end{table}

In our exploration and analysis, we operationalized 24 metrics from the CHAOSS project's library of 79 open-source software (OSS) health metrics, utilizing the data collected through the Augur project. These metrics can be grouped as: Pull Request Metrics, Issue Tracking Metrics, Contributor Metrics, Release Activity Metrics, Engagement and Popularity Metrics, and Project Dependency Metrics. The metrics described below were collected over the entire project timeline, starting from the first recorded activity on the GitHub platform to December 31, 2023. 

\subsubsection{Pull Request Metrics}

The pull request metrics feature set includes several key attributes that capture various aspects of development activity within GitHub repositories. These features are \textbf{Pull Request Count} (Pr\_count), representing the frequency of pull requests as an indicator of development activity; \textbf{Pull Request Total Files} (Pr\_total\_files), which quantifies the number of files changed across all pull requests, providing insight into the scope of modifications; \textbf{Pull Request Average Commits} (Pr\_average\_commits) and \textbf{Pull Request Total Commits} (Pr\_total\_commits), which measure the typical and cumulative number of commits per pull request, respectively, highlighting contribution granularity; \textbf{Pull Request Total Comments} (Pr\_total\_comments), reflecting the level of collaboration and peer review; and \textbf{Pull Request Review Duration} (Pr\_review\_duration\_in\_hours), which measures responsiveness in the review process. These features provide a comprehensive view of pull request dynamics, collaboration, and the development workflow.

\subsubsection{Issue Tracking Metrics}

The issue-tracking metrics feature set includes several key attributes that provide insights into managing and resolving issues within GitHub repositories. These features are: \textbf{Total Issue Duration} (Total\_issue\_duration), which measures the cumulative time issues remain open, reflecting the efficiency of issue resolution; \textbf{Average Comment Count Per Issue} (Avg\_comment\_count\_issue) and \textbf{Total Comment Count for Issues} (Total\_comment\_count\_issue), which represent the level of discussion and collaboration occurring around issues, indicating community engagement; and \textbf{Average Time to First Response Per Issue} (Avg\_ttfr\_hours), which captures the responsiveness of contributors to newly reported issues. Together, these metrics help quantify the effectiveness of issue management, community involvement, and communication within a project.

\subsubsection{Contributor Metrics}

The contributor metrics feature set provides insights into community engagement and collaboration within GitHub repositories. Key metrics include \textbf{Contributor Count} (Contributor\_count), which quantifies the total number of unique contributors over a specified period, indicating the overall involvement and diversity within a project. \textbf{New Contributor Count} (New\_contributor\_count) measures the influx of first-time contributors, offering insights into the project's capacity to attract new members and promote inclusivity. Additionally, Committer Count represents the total number of individuals whose commits have been accepted into the repository, reflecting direct code contributions. Collectively, these metrics highlight the community dynamics, sustainability, and collaborative nature of software development projects. Additionally, \textbf{Bus Factor} (Bus\_factor) measures the risk associated with the concentration of critical knowledge among a few contributors, indicating the resilience of the project to unexpected team member departures. These metrics help understand the project's architectural complexity, potential maintenance challenges, and resilience against disruptions.

\subsubsection{Release Activity Metrics}

The release activities metrics capture essential aspects of the software development lifecycle in GitHub repositories. \textbf{Release Count} (Release\_count) indicates the number of official releases made, reflecting the project's maturity and software delivery cadence. This feature helps assess the progress in software development.

\subsubsection{Engagement and Popularity Metrics}

The engagement and popularity metrics provide insights into the level of interest and community activity surrounding a GitHub repository. Key metrics include \textbf{Fork Count} (Fork\_count), which represents the number of times a repository has been forked, reflecting the extent to which others are interested in experimenting with or contributing to the project. \textbf{Watchers Count} (Watchers\_count) measures the number of users who have subscribed to updates about the repository, indicating their ongoing interest in its development. \textbf{Stars Count} (Stars\_count) quantifies the number of users who have "starred" the repository as a form of appreciation or endorsement. These metrics highlight the repository's community engagement, interest, and overall popularity within the broader open-source ecosystem.

\subsubsection{Project Dependency Metrics}

The project complexity and dependencies metrics provide insights into a GitHub repository's structural complexity and external dependencies. Key features include \textbf{Dependency Count} (Dependency\_count), which represents the total number of external libraries, frameworks, or packages that the project relies on, reflecting the complexity and maintainability challenges associated with managing these dependencies. 

More detailed descriptions for each involved feature can be found in Table \ref{tab:feature-description} within the Appendix section.

\subsection{Methods}

\subsubsection{Data Preprocessing}

In the data preprocessing stage, we conducted several steps to prepare the dataset for training the machine learning classification model. First, we employed a simple imputation strategy to handle missing values in the feature set by filling them with zeros. This approach maintained the dimensionality of the feature set while providing a consistent value, thereby minimizing bias in the subsequent analysis. Our use of standardized metrics that can be retrieved and calculated for any software project using an open source platform like GitHub ensures that the use of zero values to fill in where no data exists for a time period holds construct validity \citep{cherryholmes_construct_1988}. 

We used the Isolation Forest \citep{liu2008isolation} algorithm to identify and remove extreme outliers from the feature set to address data quality issues arising from outliers. Isolation Forest is an unsupervised machine learning algorithm well-suited for anomaly detection, particularly for datasets with high-dimensional features where defining specific outlier thresholds is challenging. The algorithm isolates anomalies by creating random splits across the features, effectively identifying data points that deviate significantly from the majority. Given that our feature set consists of over 20 numerical features, we selected Isolation Forest due to its effectiveness with high-dimensional numerical data and ability to detect complex patterns. The phenomena of open source software include many projects whose metrics and measurement will lie outside an overwhelming majority of other projects, the Linux Kernel being the most well-known among them \citep{yu_categorization_2004}. 

Given the differing sample sizes and characteristics of the three classes ("graduated," "incubating," and "sandbox"), we configured distinct contamination levels for each class to reflect their respective distributions accurately. Specifically, we set the contamination level to 0.01 for the "graduated" class, 0.05 for the "incubating" class, and 0.10 for the "sandbox" class. This tailored approach to anomaly detection ensured the removal of data points that exhibited extreme deviations within each class. Following outlier exclusion, the sample sizes were adjusted to 22 for the "graduation" class, 30 for the "incubating" class, and 101 for the "sandbox" class, resulting in a total of 153 samples retained for model training. This reduction from the initial 165 samples improved data reliability by mitigating the influence of outliers, ultimately contributing to a more robust classification model. The expected contamination range between each class demonstrates face validity with the underlying phenomena; one expects limited contamination in the most well-established (graduated) projects and an incrementally more significant amount in the more experimental projects classified as "sandbox".  

\subsubsection{Model Training, Validation and Interpretation}

\paragraph{Training and Testing Dataset Splitting}

We first split the dataset into training and testing subsets for the model training and validation process. Specifically, 80\% of the samples were allocated to the training dataset, while the remaining 20\% were designated for testing. To ensure that the class distributions were adequately represented in the training and testing datasets, we employed a stratified sampling technique \citep{liberty2016stratified}. Given the differing sample sizes across the three classes, this approach was critical, helping to maintain proportional representation and mitigate potential biases that could arise from uneven class distributions. 

\paragraph{Model Selection and Rationale}

We began by conducting an exploratory analysis of our feature set. Box plot examinations revealed the presence of outliers and skewness in each feature, while the Shapiro-Wilk test \citep{SHAPIROWILK} indicated that none of the features were normally distributed. Furthermore, Box's M test \citep{BoxMTest} showed that the covariance matrices differed significantly across classes, implying that the data do not adhere to uniform variance-covariance assumptions. In addition, Spearman correlation tests \citep{SpearmanCor} and partial dependence plots \citep{greenwell2017pdp} highlighted that several features were highly correlated with one another and that non-linear relationships existed between the features and the target variable (i.e., the three lifecycle groups). Our feature set also contains both continuous and discrete numeric subtypes, necessitating careful preprocessing to handle the diverse data distributions. 

Based on these observations and a careful review of model suitability, we selected the following algorithms to classify the groups: Support Vector Machines with a Radial Basis Function kernel (SVM) \citep{scholkopf1997comparing}, Decision Trees \citep{song2015decision}, Random Forests \citep{breiman2001random}, and Gradient Boosting \citep{friedman2001greedy}. These methods are well suited to (1) complex, non-linear feature-outcome relationships, (2) mixed numeric feature subtypes, (3) data prone to noise and outliers, (4) intricate feature interactions, (5) data that deviate from normal distributions, and (6) no uniform variance--covariance across different classes. Although larger sample sizes typically benefit these algorithms, prior studies \citep{CHI20081793,chang2011mining,ijerph17196997,ijerph18168530} have demonstrated their efficacy even in small-sample settings.

\paragraph{Preprocessing for Modeling}

To ensure that the models with sensitivity to feature scaling performed optimally, we applied the StandardScaler \citep{ahsan2021effect} transformation to the feature set before training the SVM model. This scaling process helped standardize the feature values, ensuring each feature contributed equally to the model training. On the other hand, no scaling techniques, such as StandardScaler, were applied before training the Decision Tree, Random Forest, and Gradient Boosting models, as these tree-based algorithms are inherently insensitive to the scale of the input features. This combination of models and preprocessing techniques provided a comprehensive assessment of various machine learning approaches, both linear and non-linear, for the classification task.

Additionally, given the differences in the number of samples across the three classes, we employed the Synthetic Minority Over-sampling Technique (SMOTE) combined with Tomek Link to address the class imbalance \citep{swana2022tomek}. SMOTE generated synthetic samples for the minority classes, thereby balancing the class distribution. Tomek Link was then applied to remove overlapping instances between classes, enhancing the separation between different class boundaries. This combined resampling method helped mitigate issues arising from unbalanced class numbers, improving the performance and generalizability of the machine learning models by ensuring a more representative training dataset.

\paragraph{Model Training Process}

To ensure each classifier was optimally tuned and rigorously evaluated, we began by defining search ranges for each algorithm’s primary hyperparameters based on preliminary experimentation:

\begin{itemize}
    \item \textbf{SVM}: we searched over a grid of the \textit{penalty parameter} with the range of [0.001, 0.001, 0.1, 1, 10, 100] and the \textit{kernel coefficient} with the range of [0.001, 0.01, 0.1, 1, 10]. 
    \item \textbf{Decision Tree}: we varied the \textit{max depth} with the range of [3, 5, 7, 10, 15], \textit{minimum samples split} with the range of [2, 5, 10], the \textit{minimum samples per leaf} with the range of [1, 2, 5, 10], the \textit{max leaf nodes} with the range of [None, 10, 20, 50], the \textit{cost-complexity pruning} with the range of [0, 0.0001, 0.001, 0.01].
    \item \textbf{Random Forest}: We tuned the \textit{number of trees} with the range of [50, 100, 200], the \textit{maximum depth} with the range of [5, 10, 15], and the \textit{minimum samples per leaf} with the range of [1, 2, 5].
    \item \textbf{Gradient Boosting}: We searched over the \textit{learning rate} with the range of [0.001, 0.01, 0.1], the \textit{maximum depth} with the range of [3, 5, 7], and the \textit{number of boosting stages} with the range of [50, 100, 200].
\end{itemize}

We performed hyperparameter tuning via a grid search \citep{Belete02092022} within each cross-validation fold, selecting the best parameter combination based on the average accuracy. By systematically exploring these hyperparameters, we minimized the risk of overfitting.

To obtain reliable performance estimates and address the multi-class imbalance, we adopted 10-fold stratified cross-validation \citep{StratifiedTenFold} repeated 10 times. Specifically: 1) The dataset was split into 10 stratified folds, ensuring each fold reflected the overall class distribution. 2) For each iteration, the model was trained on nine folds and validated on the remaining fold, with this process repeated 10 times so that every fold served as the validation set exactly once. 3) The entire 10-fold routine was then repeated 10 times with different random seeds, yielding 100 total runs per model. (4) The final performance metric was computed by averaging results over all runs, providing a robust estimate and reducing the influence of any single split.

Once the optimal hyperparameters were identified from the repeated cross-validation, we retrained each model on the entire training set with its best configuration. This step produced a final, fully tuned model that leveraged all available training data.

\paragraph{Feature Selection}

To improve the efficiency and performance of the classification models, we employed the forward sequential feature selection method for feature selection \citep{aha1995comparative}. Forward sequential feature selection is an iterative technique that starts with an empty set of features and progressively adds the most significant feature at each step, based on a predefined evaluation criterion. This approach aims to identify the optimal subset of features that contribute the most to model performance, thereby reducing the risk of overfitting and improving computational efficiency. By selecting only the most relevant features, we ensured that the models were trained on a more informative and less redundant feature set, enhancing their predictive power and interpretability.

\paragraph{Model Evaluation and Finalization}

After training each classifier with its best-performing subset of features, we evaluated performance on a separate test set that was held out during hyperparameter tuning and training. The following metrics were used:

\begin{itemize}
    \item \textbf{Precision} indicates how many of the instances predicted for a given class actually belong to that class. 
    \item \textbf{Recall} measures how many of the true instances of a class were correctly predicted as such.
    \item \textbf{F1-score} is the harmonic mean of precision and recall, providing a single measure that balances both.
    \item \textbf{Accuracy} gives the proportion of correct predictions across all classes.
    \item \textbf{Macro-averaged metrics (precision, recall, F1)} treat each class equally, making them suitable for cases with class imbalance.
    \item \textbf{Weighted-averaged metrics} adjust the contribution of each class according to its support size, providing a metric closer to overall accuracy but still sensitive to per-class performance.
\end{itemize}

To select the most suitable model, we compared the accuracy, macro-averaged F1-score, and weighted-averaged F1-score of each final classifier on the test set. Both Decision Tree and Gradient Boosting demonstrated the highest performance among the models evaluated. However, we ultimately chose the Decision Tree as our final model due to its computational efficiency. This selection balances predictive accuracy and lower computational overhead, making it particularly practical in environments with limited computational resources. Table~\ref{tab:classification_report} summarizes the Decision Tree’s final performance.

\paragraph{Model Result Interpretation}

We calculated the feature importance scores to interpret the results of the final model and gain insights into the contributions of each selected feature. The feature importance scores quantified the extent to which each feature chosen contributed to the overall classification task, allowing us to identify the most influential features in the model \citep{liu1994importance}. In addition to calculating feature importance, we also generated Ridgeline plots \citep{wilke2022package} for each of the selected features. These plots visually illustrated how the distributions of each feature varied across the three classes, providing a deeper understanding of the distinguishing characteristics of each class. This combination of quantitative and visual analyses facilitated a comprehensive interpretation of the model, enhancing its explainability and aiding in understanding the underlying patterns driving the classification results.

\subsection{Engaged Field Research}
An 8-year engaged field research study of the CHAOSS project within the Linux Foundation informs our empirical work. The type of engaged field research we do is specified through soft systems methodology \citep{checkland2006learning, reynolds_soft_2010} and trace ethnography \citep{geiger_trace_2011, goggins_context_2013, goggins_creating_2013}. We are transparent about our roles with our informants \citep{etherington_ethical_2007}. To make certain that we present our research with resonance \citep{haraway_situated_1988}, and ensure our descriptions of our field sites are not from the perspective of an "outside place" \citep{taylor_out_2011}, we incorporate our reporting of research with utility back the communities we engage with \citep{haraway_situated_1988, barad_meeting_2007}. 

\section{Findings}

\subsection{Findings for the First Research Question}

To answer the first research question: \textit{What aspects of open-source software project activity, including community engagement, responsiveness, and complexity, are the most significant indicators of a project’s lifecycle stage?} We present our findings and the corresponding findings of the classification model's performance and selected features' importance rates, as shown below.  

\subsubsection{Model Performance}

\begin{table}[ht]
\centering
\begin{tabular}{lcccc}
\toprule
\textbf{Class} & \textbf{Precision} & \textbf{Recall} & \textbf{F1-score} & \textbf{Support} \\
\midrule
grads & 1.00 & 0.75 & 0.86 & 4 \\
incubating & 0.83 & 0.83 & 0.83 & 6 \\
sandbox & 0.91 & 0.95 & 0.93 & 21 \\
\midrule
\textbf{Accuracy} & & & 0.90 & 31 \\
\textbf{Macro avg} & 0.91 & 0.85 & 0.87 & 31 \\
\textbf{Weighted avg} & 0.91 & 0.90 & 0.90 & 31 \\
\bottomrule
\end{tabular}
\caption{Classification Report}
\label{tab:classification_report}
\end{table}

The classification performance of our Decision Tree model, as reported in Table \ref{tab:classification_report}, demonstrates high accuracy in predicting the lifecycle stage of open-source software projects across the three classifications: "sandbox," "incubating," and "graduated." Specifically, the model achieves an overall accuracy of 90\%, indicating its effectiveness in distinguishing between the stages. The precision, recall, and F1-score metrics offer deeper insights into the model’s performance for each class. For "sandbox" projects, the model achieves a high recall of 0.95 and an F1-score of 0.93, suggesting that the model can consistently identify sandbox projects with minimal error. Similarly, "incubating" projects yield an F1-score of 0.83, with balanced precision and recall. This indicates that while the model is generally effective at identifying these projects, there is room for improvement in capturing subtle distinctions.

For "graduated" projects, the model reaches perfect precision at 1.00 but with a recall of 0.75, which implies that while all identified graduated projects are correctly placed, the model occasionally misclassifies graduated projects into other stages. The macro and weighted averages, with F1-scores of 0.87 and 0.90, respectively, further support the overall robustness of the model in predicting project stages. These metrics reflect the model's ability to provide reliable classifications across diverse project types, underscoring the efficacy of the Decision Tree model in supporting our lifecycle classification task.

\subsubsection{Feature Importance}

Table \ref{tab:feature_importance} presents the most influential features the Decision Tree model uses to distinguish among the lifecycle stages of open-source software projects. The feature \textit{new\_contributor\_count} stands out with the highest importance score of 0.561794, suggesting that the influx of new contributors plays a crucial role in determining a project's progression within the lifecycle. This high feature importance aligns with expectations, as projects with more contributors, especially new ones, are likely at more advanced stages, benefiting from broader interest and engagement from the developer community.

The following set of features, highlighted in orange and green, includes \textit{stars\_count} (0.073238), \textit{pr\_average\_commits} (0.064346), \textit{dependency\_count} (0.057202), and \textit{avg\_ttfr\_hours} (0.054981). These features likely capture key signals of project maturity: \textit{stars\_count} reflects community interest, while \textit{pr\_average\_commits} and \textit{dependency\_count} suggest project complexity and dependence on other tools, which are indicators of stability and readiness for adoption. The \textit{avg\_ttfr\_hours} (average time to first response) likely signals responsiveness, an important quality as projects advance toward a ``graduated'' stage.

\begin{table}[H]
\centering
\begin{minipage}{0.5\textwidth}
\centering
\begin{tabular}{lr}
\hline
Feature & Importance \\
\hline
\rowcolor{cb_yellow!20} \textbf{new\_contributor\_count} & \textbf{0.561794} \\
\rowcolor{cb_orange!20} stars\_count & 0.073238 \\
\rowcolor{cb_orange!20} pr\_average\_commits & 0.064346 \\
\rowcolor{cb_green!20} dependency\_count & 0.057202 \\
\rowcolor{cb_green!20} avg\_ttfr\_hours & 0.054981 \\
\rowcolor{cb_blue!20} pr\_total\_files & 0.048365 \\
\rowcolor{cb_blue!20} average\_comment\_count & 0.038707 \\
\rowcolor{cb_blue!20} commits & 0.037535 \\
\rowcolor{cb_blue!20} pr\_total\_comments & 0.031660 \\
\rowcolor{cb_purple!20} bus\_factor & 0.016086 \\
\rowcolor{cb_purple!20} comments\_per\_issue & 0.016086 \\
\hline
\end{tabular}
\caption{Feature Importance Table}
\label{tab:feature_importance}
\end{minipage}%
\hfill
\begin{minipage}{0.4\textwidth}
\centering
\begin{tabular}{lc}
\hline
Color & Description \\
\hline
\rowcolor{cb_yellow!20} Yellow & Importance $>$ 0.08 \\
\rowcolor{cb_orange!20} Orange & 0.06 $<$ Importance $\leq$ 0.08 \\
\rowcolor{cb_green!20} Green & 0.04 $<$ Importance $\leq$ 0.06 \\
\rowcolor{cb_blue!20} Blue & 0.02 $<$ Importance $\leq$ 0.04 \\
\rowcolor{cb_purple!20} Light Blue & Importance $\leq$ 0.02 \\
\hline
\end{tabular}
\caption{Legend for Feature Importance Colors: The colors—yellow, orange, green, blue, and light blue—represent different ranges of feature importance scores. Light blue indicates features with the lowest importance scores, blue represents features with small importance, green represents moderately important features, orange highlights features with relatively higher importance, and yellow marks the features with the highest importance scores}
\label{tab:color_legend}
\end{minipage}
\end{table}

The model also includes several additional features of smaller importance, denoted in blue and light blue shades. These features, such as \textit{pr\_total\_files}, \textit{average\_comment\_count}, and \textit{bus\_factor}, represent various aspects of project activity and engagement. Although these features individually contribute less to the model, their presence highlights nuanced behaviors and project characteristics that collectively support the model's predictive performance. The varied importance of these features, visualized in Table~\ref{tab:color_legend}, reinforces that different aspects of project activity, from community engagement to code complexity, collectively inform the project lifecycle classification. 

\subsection{Findings for the Second Research Question}

In this section, we present findings related to our second research question: \textit{How do sociotechnical factors shape the classification boundaries, progression patterns, and interpretability of models classifying OSS project lifecycle stages?}

Our analysis identified eleven features significantly influencing the model's classification of OSS lifecycle stages. Each of these features carries distinct sociotechnical implications. To answer the second research question, we employed ridgeline plots for the following reasons:

\begin{itemize}
    \item While feature importance metrics indicate the relative influence of each feature in the classification model, they do not capture the actual distribution of these features across lifecycle stages. Ridgeline plots address this limitation by visualizing the spread and density of each feature within the sandbox, incubating, and graduated stages, allowing for a more nuanced understanding of how each sociotechnical factor differentiates these stages.
    \item Ridgeline plots reveal natural breaks and overlaps in feature values across stages, offering insights into how the model distinguishes stages based on sociotechnical characteristics. By identifying these patterns, the ridgeline plots provide context for the model’s decision boundaries, showing where classification may be more reliable and where stage boundaries might blur.
    \item By visualizing the continuous progression of feature values across lifecycle stages, ridgeline plots illustrate how sociotechnical factors evolve as OSS projects mature. This dynamic perspective is essential for understanding how certain features reflect developmental trends within OSS projects, supporting the model’s interpretation of project maturity.
    \item Finally, ridgeline plots enhance interpretability by aligning feature distributions with real-world characteristics of OSS lifecycle stages. This alignment allows for a more intuitive understanding of each factor's role, enabling clearer insights into how specific sociotechnical factors drive distinctions between stages.
\end{itemize}

We believe that the ridgeline plots' unique insights complement traditional performance metrics and feature importance scores, enriching our understanding of how sociotechnical factors influence the classification of OSS project lifecycle stages.

In the following subsection, we detail the insights gained from the ridgeline plots for the four most influential features identified in the previous section.

\subsubsection{Four Most Influential Factors}

\begin{figure}[h! tbp]
\begin{subfigure}{0.45\textwidth}
\includegraphics[width=1\linewidth, height=4cm]{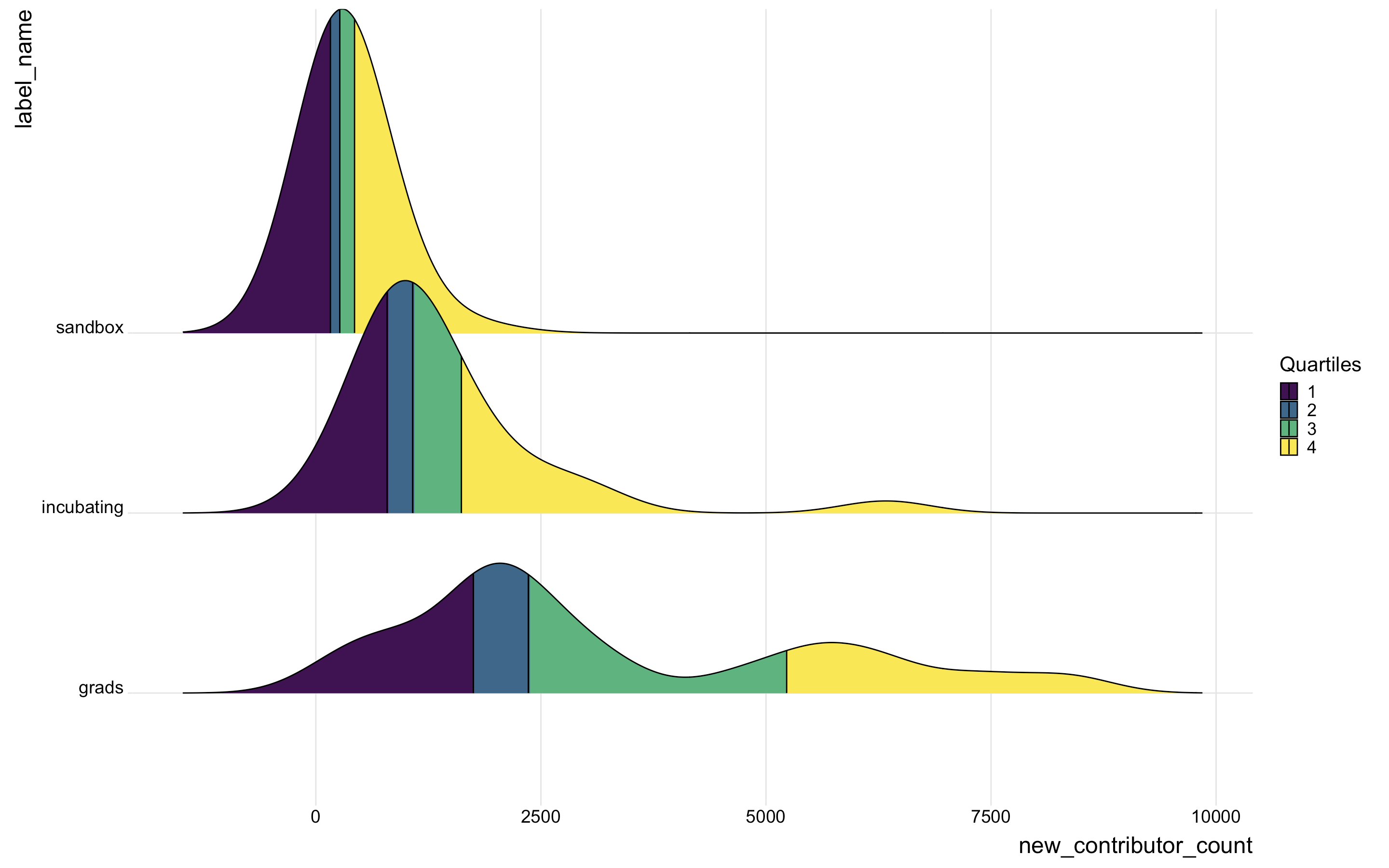}
\caption{New contributor count}
\label{fig:subim11}
\end{subfigure}
\begin{subfigure}{0.45\textwidth}
\includegraphics[width=1\linewidth, height=4cm]{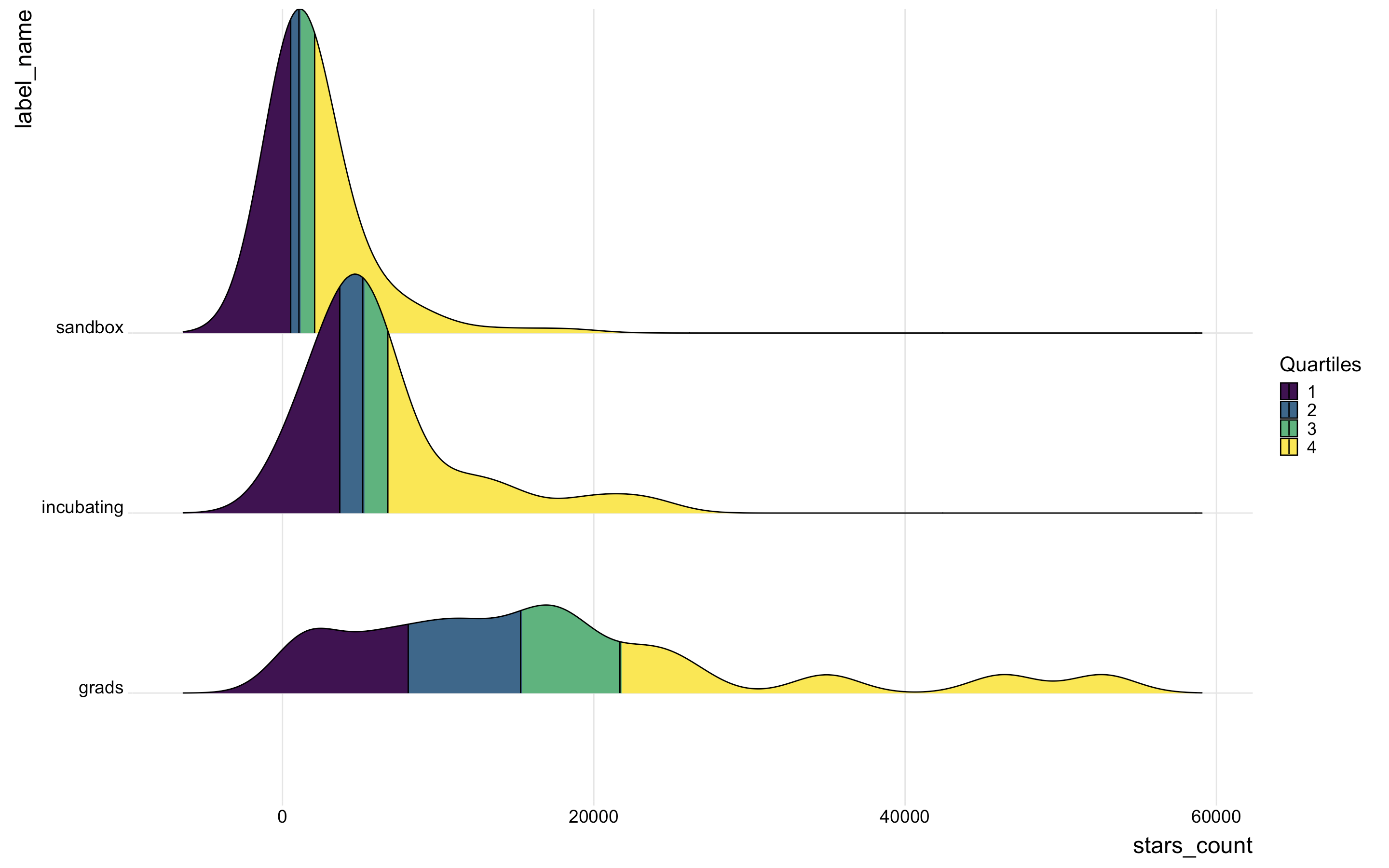}
\caption{Stars Count}
\label{fig:subim12}
\end{subfigure}

\begin{subfigure}{0.45\textwidth}
\includegraphics[width=1\linewidth, height=4cm]{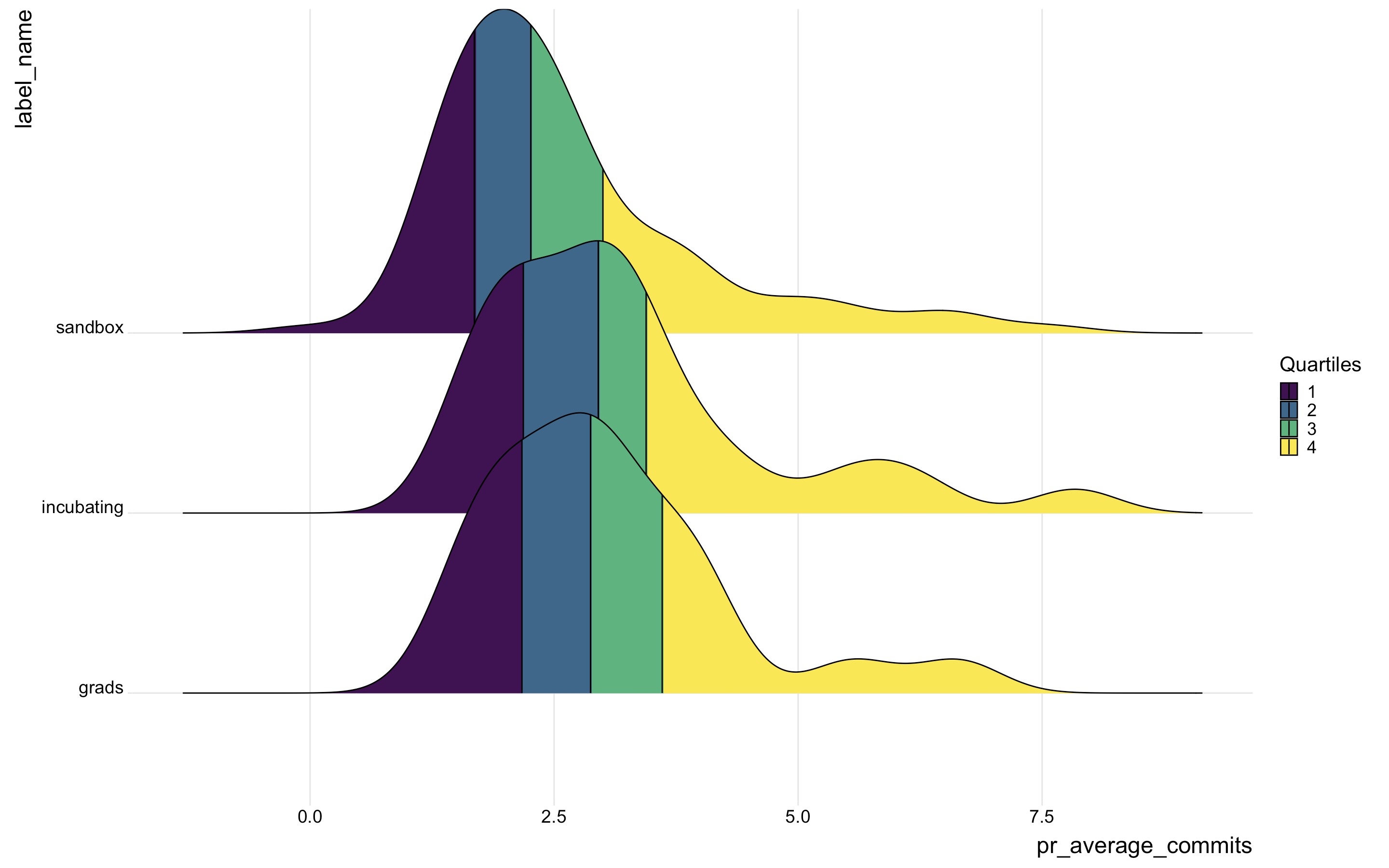}
\caption{Pull Request Average Commits}
\label{fig:subim13}
\end{subfigure}
\begin{subfigure}{0.45\textwidth}
\includegraphics[width=1\linewidth, height=4cm]{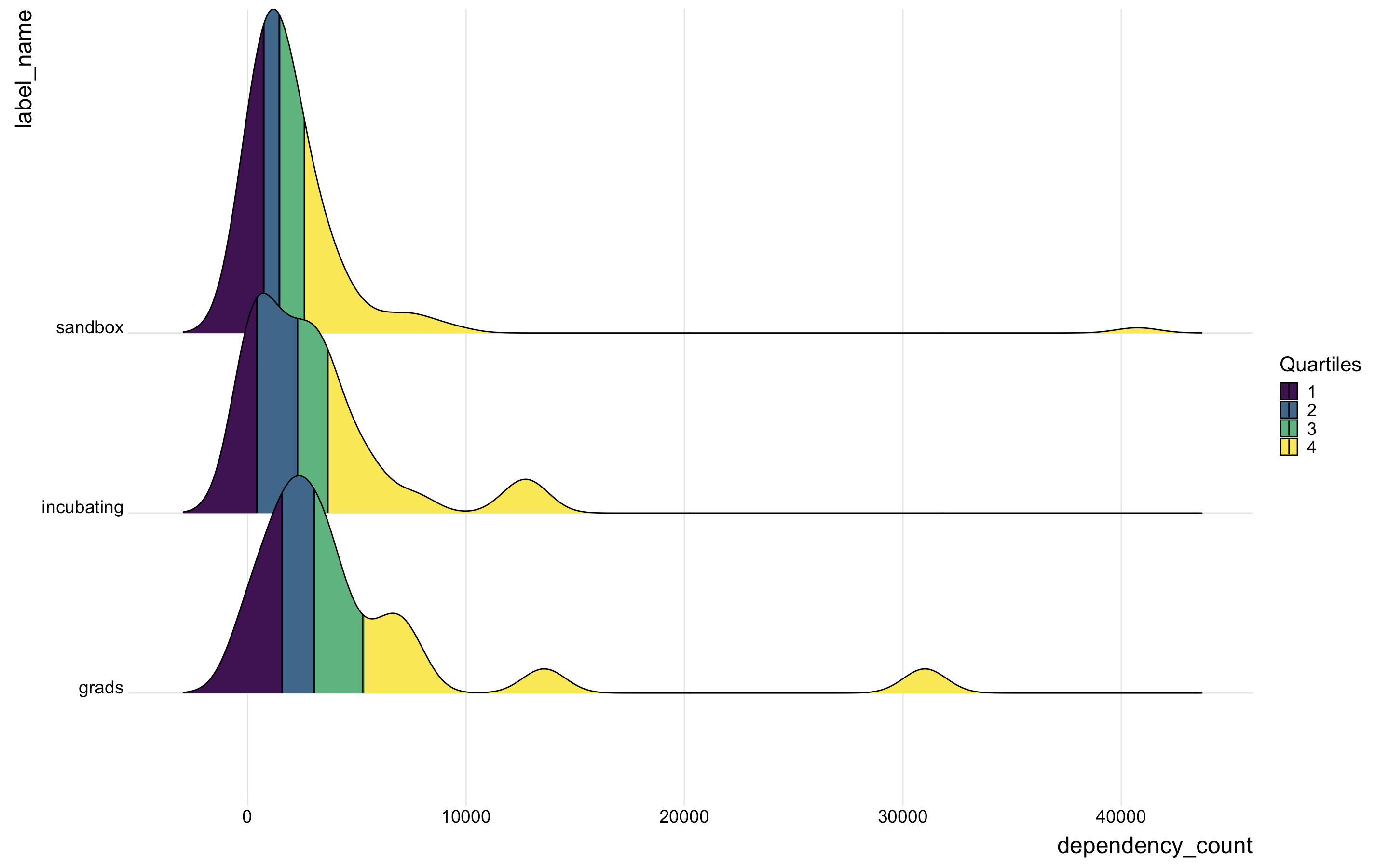}
\caption{Dependency Count}
\label{fig:subim14}
\end{subfigure}

\caption{These ridge line plots visually represent how the quartile distribution of CNCF projects for each of our model's four most significant metrics distinguishes between the three classes. In each plot, "sandbox" projects are at the top, "incubating" projects are in the middle, and "grads" are last.}
\label{fig:image1}
\end{figure}

Figure \ref{fig:image1} presents ridgeline plots for the data distribution of the four most influential features across each CNCF project lifecycle stage. These plots visually represent how these key sociotechnical factors vary by lifecycle stage and contribute to the model’s classification. 

\paragraph{New Contributor Count}

Specifically, the first ridgeline plot (Figure \ref{fig:subim11}) illustrates the distribution of \textit{New Contributor Count} across the three OSS lifecycle stages. This plot provides insight into how the number of new contributors varies across stages, highlighting how this sociotechnical factor impacts the model’s classification. Each ridge corresponds to a different stage, with sandbox projects at the top, followed by incubating projects and graduated projects at the bottom. The x-axis shows the \textit{New Contributor Count}, and colors within each ridge represent quartiles, with darker shades indicating lower counts and lighter shades indicating higher counts. This visual segmentation by quartile reveals shifts in \textit{New Contributor Count} as projects progress through lifecycle stages.

In sandbox projects, the distribution skews toward lower values, indicating relatively low new contributor activity, as represented by the darker colors concentrated on the left side of the plot. This aligns with the characteristics of sandbox projects, which are typically in early development stages and have not yet attracted substantial community engagement. The model likely uses this low \textit{New Contributor Count} as a distinguishing feature of sandbox projects. In incubating projects, the distribution shifts slightly to the right, with a broader range of counts. Although lower quartiles still appear prominently, a moderate level of new contributor activity is also evident, reflecting the growth in community involvement typical of projects in intermediate stages.

The distribution extends further to the right for graduated projects, with a greater spread across higher counts, as indicated by the lighter colors. This pattern is consistent with mature, widely adopted projects that attract significant community engagement. The model likely interprets high \textit{New Contributor Count} as a strong indicator of the graduated stage, differentiating it from earlier stages based on this key sociotechnical factor. 

\paragraph{Stars Count}

The ridgeline plot for \textit{Stars Count} Figure \ref{fig:subim12} shows the distribution of this feature across the three lifecycle stages. Stars Count reflects the level of community interest or endorsement for each project. This plot provides insights into how community recognition influences the model’s classification boundaries, progression patterns, and interpretability.

The distribution skews strongly towards lower star counts in sandbox projects, with most data concentrated in the first and second quartiles. This clustering suggests that sandbox projects generally have limited visibility and community endorsement, characteristics of early-stage projects yet to gain traction. This low \textit{Stars Count} serves as a clear signal for the model, helping it to identify sandbox projects. The distribution broadens and shifts slightly rightward for incubating projects, indicating increased community recognition. Although lower quartiles still appear prominently, higher quartiles also begin to emerge, suggesting that incubating projects attract moderate interest.

The distribution expands even further in graduated projects, with substantial representation in the higher quartiles, particularly in the third and fourth. This shift reflects the strong community endorsement and popularity typically associated with mature projects. The model likely relies on this high \textit{Stars Count} to distinguish graduated projects from earlier stages, recognizing it as an indicator of the project's stability and widespread adoption within the open-source ecosystem.

\paragraph{Pull Request Average Commits}

The ridgeline plot for \textit{PR Average Commits} Figure \ref{fig:subim13} illustrates the distribution of the average number of commits per pull request across lifecycle stages. This feature reflects the depth or complexity of individual contributions, which varies with the project's maturity.

The distribution is skewed towards lower values for sandbox projects, with most data concentrated in the first and second quartiles. This pattern suggests that pull requests in sandbox projects generally involve fewer commits, possibly reflecting simpler, exploratory changes characteristic of early-stage projects. The model likely associates this low average with the sandbox stage, using it as a signal of limited complexity in contributions. The distribution shifts rightward and broadens for incubating projects, with a notable presence in the third quartile, indicating moderate contribution complexity. This shift suggests that incubating projects attract contributors working on more involved or multifaceted changes, which is characteristic of projects in a transitional growth stage.

In graduated projects, the distribution extends further rightward, with a substantial portion of data in the fourth quartile. This reflects a higher level of complexity in pull requests, aligning with the characteristics of stable, mature projects that require comprehensive contributions to support advanced functionality or maintenance. The model leverages this pattern to identify graduated projects, associating higher complexity in pull request activity with this mature stage.

\paragraph{Dependency Count}

The ridgeline plot for \textit{Dependency Count} Figure \ref{fig:subim14} shows how dependencies vary across OSS lifecycle stages. Dependency Count represents the extent to which a project relies on external components, with higher counts often indicative of greater complexity and integration.

In sandbox projects, the distribution is skewed towards lower values, with most data concentrated in the first and second quartiles. This clustering suggests that sandbox projects generally have fewer dependencies, reflecting their early development focus on core functionality rather than extensive integrations. The model uses this low \textit{Dependency Count} to identify sandbox projects, associating fewer dependencies with projects at an early stage. For incubating projects, the distribution shifts slightly to the right and broadens, with some representation in the third and fourth quartiles. This shift indicates that incubating projects are beginning to incorporate more dependencies, signaling a progression towards greater complexity as projects mature.

In graduated projects, the distribution shows a much broader spread across higher values, with notable representation in the third and fourth quartiles. This higher dependency count reflects the complex architecture and extensive functionality typical of mature projects that rely on multiple external components. The model likely uses this pattern to classify graduated projects, identifying higher \textit{Dependency Count} as a stability and advanced functionality marker.


Additionally, Appendix A includes the remaining seven features or sociotechnical factors.

\subsubsection{Sociotechnical Nature of Four Most Influential Factors}

To answer this research question, this section describes the four most influential factors in sociotechnical terms. Appendix A includes the remaining seven factors. For the first research question, we centered our empirical presentation on accepted measurements of model quality. Such measurements do not facilitate understanding the differences in data distribution within and across the factors. We use ridge plots and field notes from our engaged field research data to describe how factors impact accuracy and reliability statistically and sociotechnically. First, we describe the importance of each metric in more detail. Second, we frame their sociotechnical operationalization by contrasting the distinct social and technical characteristics. 

The factors we modeled to classify projects according to their lifecycle location all arise from the work of over 4,000 contributors to the CHAOSS project over the past eight years. That project calls these factors "metrics," and each one is defined to help open source projects understand and articulate their health and sustainability. In practice, metrics do not stand alone or provide convenient material for management dashboards inside technology corporations that contribute to the Linux Foundation and other open source projects. Instead, these metrics, which we applied to a classification problem, arise from a combination of open source projects' social and political needs of open source projects, which intersect with some desire to understand a technical component. 

New project contributors are recognized as necessary for open source project sustainability. Welcoming newcomers is a practice honed within the CHAOSS project and spread beyond its boundaries. Recognition of the importance of newcomers arises from experiences within open source where projects were not very good at it, and those projects learned they needed to become better at this to sustain themselves. The social practices for welcoming newcomers were developed to ensure that the technology managed by any given open source project would be available for companies and other organizations that rely on it. This is a social and political factor with technological implications. 

Stars, in contrast, are lightweight indicators of the awareness others have of a given project. A star may or may not indicate popularity. In contrast, it means attention that can be leveraged to recruit new contributors, corporate participation, and other resources of value for sustaining an open source project. Open source projects cannot live on or eat their stars, but we do know that projects with more stars have more opportunities to get the things they need. 

In contrast with the first two factors, pull request average commits are primarily technical information that strongly signals how an open source project is orchestrated. They signal the approximate amount of new code suggested in pull requests. Because open source software maintainers carry a heavy workload, there is a size of a pull request that, for practical purposes, is too big to merge. The more the software changes, the more complex the task for the person reviewing the pull request, and the less likely it is that they will have time to do it. This technical measure of scale, which influences social and political considerations of limited resources, is a strong factor in the model presented in this paper and has face validity with our field site. 

Finally, dependency count is a CHAOSS metric that has arisen in the past three years as a response to concerns expressed by the White House Office of Science and Technology Policy about the security of the "software supply chain." Every dependency holds two potentials: reducing work because the project team does not need to write that code and introducing risk because the project team does not control that code. 

\section{Discussion}

This paper built and demonstrated the validity of a computational model for differentiating open source projects by their software lifecycle positions. As complex sociotechnical systems, each open source project is difficult to classify or put in a grouping with other projects. The importance of understanding how to evaluate open source projects and decide which ones to use means that this beginning will lead to open source software project classifications using different dimensions and goals in the future. Lifecycle classification offers immediate, coarse-grained utility, though.  

Sometimes conflicting purposes may exist for classifying open source projects and how these competing interests can advance and impede our progress toward a comprehensive understanding of how open source people, projects, and technology are situated in larger ecosystems. Empirically, this paper illustrates some of the specific challenges of aligning the analysis of electronic trace data with our understanding of open source. This paper's findings may help pave the way forward and advance our scientific and practical knowledge of open source through candidate techniques for helping individuals, groups, organizations, and the societies they affect find their way across the OSS Ocean. Future work could be centered on developing wayfinding \citep{bolici_stigmergic_2016, storey_shared_2006} techniques for open source and offer a discussion about the potential for identifying and curating a set of open source genres \citep{devitt_writing_2004} to act as a "first filter" to sample and understand open source phenomena and their effects.

Classification is a form of grouping. Reciprocal and dynamic groupings are taken up in other fields as genres that constitute the situation, culture, and universe of other genres \citep{devitt_generalizing_1993, devitt_writing_2004}. Genre analysis recognizes that classification's fatal flaw is its inability to represent the different purposes and intents involved in creating a classification system and a failure to realize that the phenomena being classified are undergoing steady change. Approaching the similarity of open source projects as genres could allow us to create groupings that are useful for a purpose at a point in time without constraining the utility of our efforts with open-source-able goals. Tentatively situating open source projects in a collection of genres could enable our study of open source ecosystems to proceed in a manner that advances our scientific understanding of these new kinds of organizations while avoiding the pitfalls of convenience sampling.

\bibliographystyle{ACM-Reference-Format}
\bibliography{main}

\pagebreak
\appendix

\section{Appendix}

\subsection{Detailed Descriptions for Each Involved Feature}

\begin{longtable}{p{3cm}|p{10cm}}
\hline
\textbf{Feature   Name} &
  \textbf{Illustration} \\ \hline
\endfirsthead
\endhead
Pr\_count &
  \textbf{Pull Request Count}: This feature represents the total number of pull requests (PRs)   associated with a GitHub repository within a specified time frame. A pull request is a method of submitting contributions to a project, where a developer requests to merge code changes from one branch into another. The pull request count quantifies the frequency of proposed code changes,   reflecting the project's level of development activity, collaboration, and community engagement. \\ \hline
Pr\_total\_files &
  \textbf{Pull Request Total Files}: This feature represents the cumulative number of files modified, added, or deleted across all pull requests associated with a   GitHub repository or user within a specified time frame. It quantifies the total scope of file-level changes proposed through pull requests, providing insight into the magnitude of contributions and the extent of codebase modifications. \\ \hline
Pr\_average \\ \_commits &
  \textbf{Pull Request Average Commits}: This feature represents the average number of commits made per pull request within a GitHub repository over a specified time frame. It quantifies how many individual commits are typically included in each pull request, offering insights into the granularity of code changes and the development practices of contributors. \\ \hline
Pr\_total\_commits &
  \textbf{Pull Request Total Commits}: This metric represents the total number of commits associated with all pull requests within a GitHub repository. This metric aggregates the number of individual code changes proposed through pull requests, a primary mechanism for contributing to a repository in collaborative software development. The feature is calculated by listing all pull requests for one certain repository, obtaining the number of commits for each pull request, and summing the number of commits from all pull requests to get the total. \\ \hline
Pr\_total\_comments &
  \textbf{Pull Request Total Comments}: This feature represents the total number of comments made on all pull requests within a GitHub repository over a specified time frame. It encompasses all forms of commentary associated with pull requests, including general discussions, code review feedback, inline comments on specific code changes, and any follow-up questions or clarifications. This metric provides insights into the level of communication, collaboration, and peer review activities occurring within a project or by an individual contributor. \\ \hline
Pr\_review\_duration \\ 
\_in\_hours &
  \textbf{Pull Request Review Duration in Hours}: This feature measures the average time, in hours,   between when a pull request (PR) is opened and when the first substantive review action occurs. This includes the time until a reviewer's first review comment,   approval, or request for changes. The Pull Request Review   Duration captures how quickly a project or user responds to code contributions, reflecting the efficiency and responsiveness of the code review process within a GitHub repository. \\ \hline
Total\_issue \\ 
\_duration &
  \textbf{Total Issue Duration}: This feature represents the cumulative duration,   typically measured in hours or days, that issues remain open within a GitHub repository or for a user over a specified time frame. It quantifies the total time from when issues are created to when they are closed, providing insights into how efficiently a project or user addresses and resolves reported problems, feature requests, or questions. Total Issue Duration reflects the responsiveness and effectiveness of issue management processes within a project or by an individual contributor. \\ \hline
Avg\_comment
\\ \_count\_issue &
  \textbf{Average Comment Count Per Issue}: This feature represents the average number of comments made on each issue within a GitHub repository or associated with a user over a specified time frame. It quantifies the typical discussion, collaboration, and deliberation level per issue. The comments can include discussions, suggestions, clarifications, or any form of communication to resolve the issue. The Average Comment Count Per   Issue provides insights into the engagement level of the community, the complexity of issues, and the effectiveness of communication within a project. \\ \hline
Total\_comment \\ 
\_count\_issue &
  \textbf{Total Comment Count regarding Issue}: This feature represents the cumulative number of comments made on all issues within a GitHub repository over a specified time frame. It includes every comment associated with issues, such as discussions about bugs, feature requests, enhancements, questions, and other topics. The total Comment Count regarding the Issue provides insights into the level of interaction, collaboration, and communication within the issue-tracking system of a repository or project. \\ \hline
Total\_pr\_review \\ 
\_comments &
  \textbf{Total Pull Request Review Comments}: This feature represents the cumulative number of review comments made on all pull requests within a GitHub repository over a specified time frame. Pull request review comments are typically inline comments associated with specific lines or sections of code in a pull request. They are part of the code review process, where reviewers provide feedback, suggest improvements, or highlight issues in the proposed code changes. The Tota Pull Request Review Comments metric provides insights into the level of code review activity, the emphasis on code quality, and the collaborative dynamics within a project. \\ \hline
Bus\_factor &
  \textbf{Bus Factor}: This feature measures the risk associated with the concentration of information and capabilities within a project. Specifically, it represents the minimum number of key contributors whose departure would jeopardize the project’s continuity and success. The term originates from the hypothetical scenario of team members being unavailable due to unforeseen circumstances (e.g.,   getting “hit by a bus”). A higher Bus Factor indicates that knowledge and responsibilities are more evenly distributed among team members, reducing the risk of project disruption. In terms of how we calculate it, we first find out how many contributions each contributor has made to the project; we then arrange the contributors in descending order using the number of contributions made. We then sum up the total number of all the users'   contributions and find the 50\% of the total. Using the descending order, we add each user's contribution to a cumulative total until we reach a total that is just above 50\% of the total contributors. The number of contributors that make up that number of cumulative contributions is just above 50\% of the total contributions to the contributor absence factor. \\ \hline
Bot\_contributors \\ 
\_count &
  \textbf{Bot Contributors Count}: This feature represents a metric representing the total number of automated agents (bots) contributing to a GitHub repository over a specified time frame. Bot contributors are accounts that perform automated tasks such as code formatting, dependency updates, testing,   continuous integration, issue management, and other routine maintenance activities. These bots often operate under dedicated GitHub accounts and can significantly impact the repository's activity levels and contribution patterns.   Understanding the Bot Contributors Count helps distinguish human contributions from automated ones and provides insights into the automation level within a project. \\ \hline
Release\_count &
  \textbf{Release Count}:   This metric represents the total number of official releases made in a GitHub repository over a specified time frame. In GitHub, a release is a packaged and versioned snapshot of the codebase, often corresponding to a significant milestone such as deploying new features, bug fixes, or stable software versions. Releases are typically tagged with version numbers and may include release notes that describe changes since the last release.   The Release Count provides insights into the project’s development lifecycle,   maturity, and software delivery cadence. \\ \hline
Issue\_count &
  \textbf{Issue Count}:   This feature represents the total number of issues created in a GitHub repository over a specified time frame. An issue in GitHub is a tool for tracking tasks, enhancements, bugs, feature requests, questions, and other work items related to the project. The issue count provides insights into the project's activity level, the column of reported problems or requested features, and the engagement of the community in contributing to the project’s development through issue reporting. \\ \hline
Contributor\_count &
  \textbf{Contributor Count}: This metric represents the number of unique individuals who have contributed to a GitHub repository over a specified time frame. The   Contributor Count provides insights into the level of community involvement,   collaboration, and diversity within a project. It reflects the breadth of participation and can indicate the project’s health,   sustainability, and attractiveness to new contributors. \\ \hline
New\_contributor \\ 
\_count &
  \textbf{New Contributor Count}: This feature represents the number of unique individuals who have made their first contribution to a GitHub repository within a specified time frame. The New Contributor Count provides insights into the project’s ability to attract fresh contributors, the inclusivity of the community, and the effectiveness of outreach and onboarding practices. It reflects the influx of new talent and perspectives, which can drive innovation and enhance the project’s long-term sustainability. \\ \hline
Dependency\_count &
  \textbf{Dependency Count}: This metric indicates the total number of external libraries,   frameworks, or packages a GitHub repository relies upon. Dependencies are essential components a project uses to function correctly, providing additional functionality without writing code from scratch. The   Dependency Count includes three types of dependencies: 1) Direct Dependency: These are first-order dependencies declared in the source code and/or package manager configuration (e.g., requirements.txt, Gemfile, etc.) 2) Transitive   Dependency: These are indirect dependencies, that is, dependencies beyond first-order dependencies, also referred to as nested or second-order dependencies. For example, project A under evaluation depends on project B,   and project B depends on Project C. For project A, project C is a transitive dependency. 3) Circular Dependency: These are dependencies that, if traced,   eventually lead back to themselves. In systems that allow circular dependencies, we assume that a given dependency is only counted once in this case. This metric provides insights into the project’s complexity,   maintainability, security risks, and the extent to which it leverages external codes. \\ \hline
Avg\_ttfr\_hours &
  \textbf{Average Time to First Response Per Issue}: This metric measures the average duration between creating an issue on GitHub and the first interaction from a project maintainer or contributor. This first response can be a comment, an assignment of labels, or any acknowledgement that indicates the project team has noticed the issue. To calculate this feature, you take all the time intervals between when issues or pull requests are opened and when the first response occurs, sum them up, and then divide by the total number of issues or pull requests analyzed. \\ \hline
Fork\_count &
  \textbf{Fork Count}:   This metric represents the total number of times users have forked a GitHub repository. In the context of GitHub, forking a repository creates a personal copy of someone else’s project in your own GitHub account. This allows developers to experiment with changes without affecting the original codebase and is often the first step towards contributing to a project through pull requests. This cumulative number indicates how many times a repository has been forked. \\ \hline
Watchers\_count &
  \textbf{Watchers Count}: It is a metric on GitHub that represents the number of users subscribed to receive notifications about a repository’s activities.   When a user watches a repository, they opt to stay informed about updates such as issues, pull requests, comments, and releases. This allows them to keep track of the project’s development and participate more actively if they choose. This is the total number of GitHub users who have chosen to watch the repository. \\ \hline
Stars\_count &
  \textbf{Stars Count}:   this metric represents the total number of users who have   “starred” a repository. Starring a repository allows users to bookmark projects they find interesting, useful, or worth following. It serves as a form of endorsement or appreciation. It makes it easier for users to keep track of repositories they might want to revisit or contribute to in the future. Stars Count is the cumulative number of stars a repository has received from GitHub users. The number of stars of a GitHub repository can be seen as a proxy of its popularity \textbackslash{}cite\{borgespopularity2016\}. A high number of stars on a project can indicate that the project is widely used or appreciated in the ecosystem. GitHub often uses stars to rank projects in search results and on trending pages, increasing the visibility of popular projects to users and potential new contributors. The more a project is seen and used, the more it affects where it will be in the lifecycle. \\ \hline
Commits\_issue &
  \textbf{Total Commits for All Issues}: This feature refers to the cumulative number of commits to resolving all issues within a GitHub repository. This feature aims to capture the extent of code changes made specifically to address reported issues, encompassing bug fixes, feature implementations, enhancements, and other modifications linked to issue tracking. \\ \hline
Committer\_count &
  \textbf{Committer Count}: The "Committer Count" is a metric that represents the total number of unique individuals who have made commits to a GitHub repository.   A committer has contributed code changes that have been accepted into the repository's codebase. This metric reflects the breadth of direct code contributions and offers insights into the project's collaborative nature. \\ \hline
Comments\_per \\ 
\_issue &
  \textbf{Average Comments Per Issue}: This metric measures the average number of comments on each issue within a GitHub repository. This metric provides insights into the level of discussion, collaboration, and engagement around reported issues, including bug reports, feature requests, or general questions related to the project. This metric is calculated by dividing the total number of comments on all issues by the total number of issues in the repository. \\ \hline
\caption{Detailed description for all 24 features}
\label{tab:feature-description}\\
\end{longtable}

\subsection{Other Ridgeline Plots of Selected Features}

\begin{figure}[h]
\begin{subfigure}{0.45\textwidth}
\includegraphics[width=1\linewidth, height=4cm]{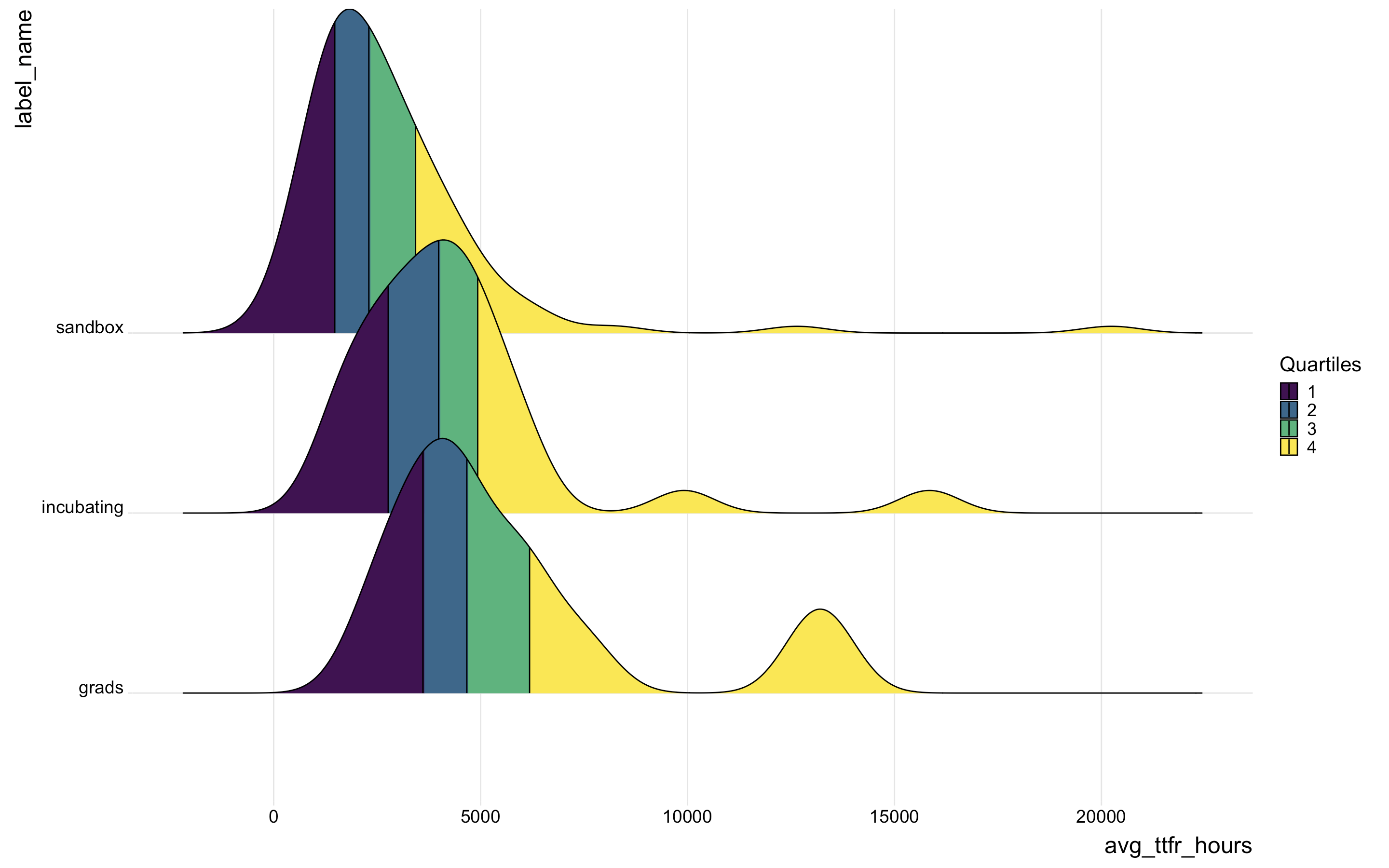}
\caption{Average time to first response hours}
\label{fig:subim1}
\end{subfigure}
\begin{subfigure}{0.45\textwidth}
\includegraphics[width=1\linewidth, height=4cm]{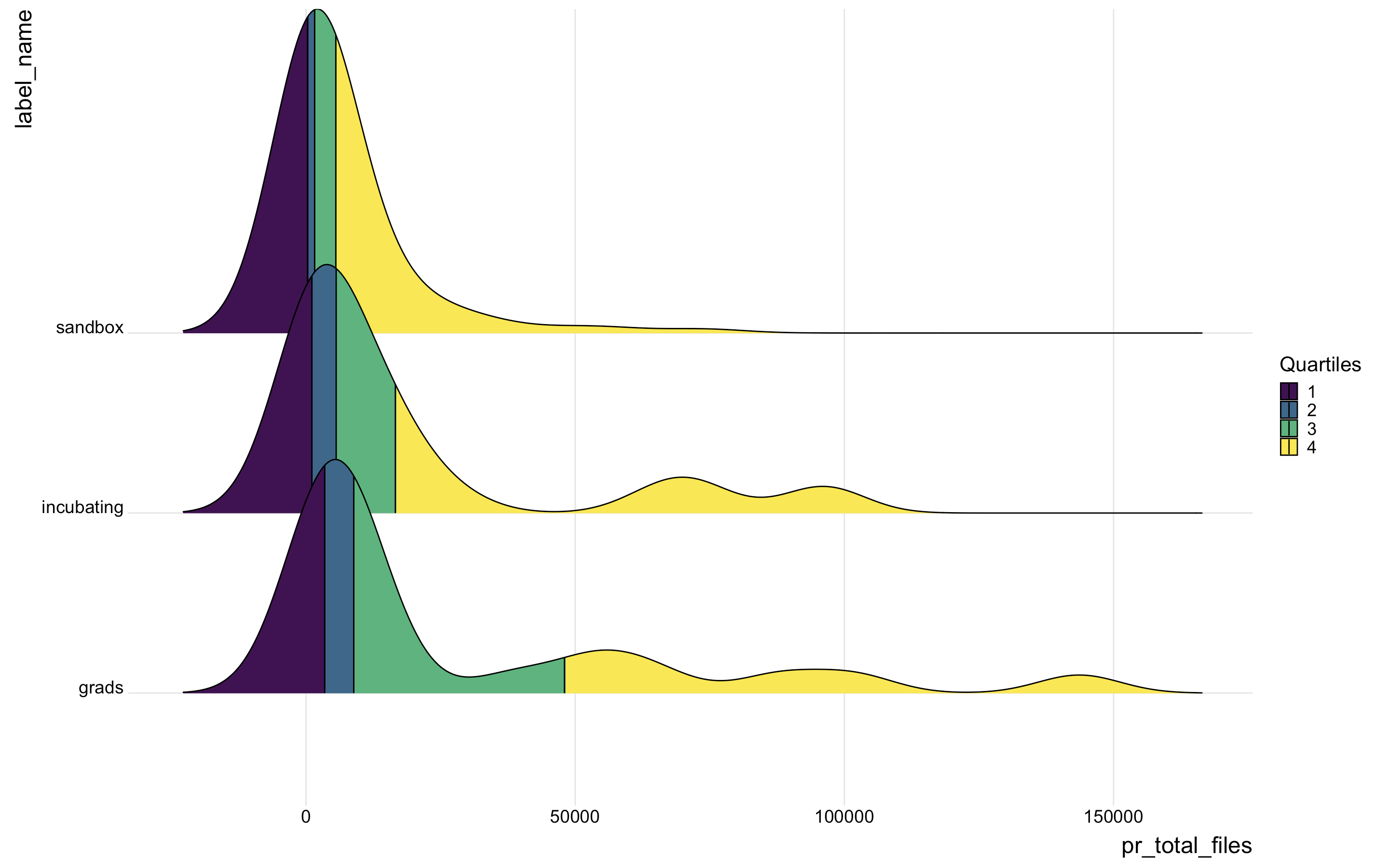}
\caption{Pull request total files}
\label{fig:subim2}
\end{subfigure}

\begin{subfigure}{0.45\textwidth}
\includegraphics[width=1\linewidth, height=4cm]{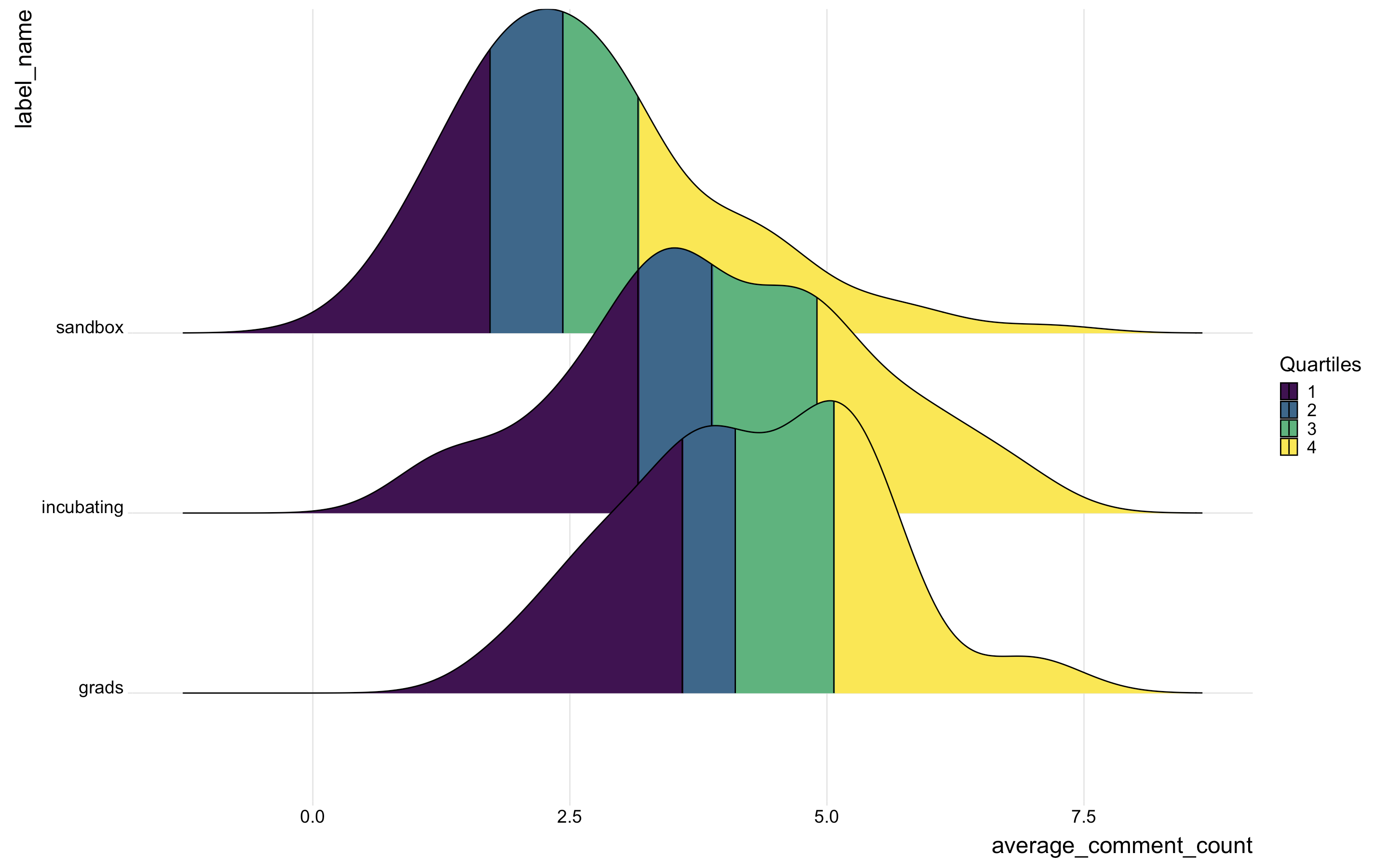} 
\caption{Average comment count}
\label{fig:subim3}
\end{subfigure}
\begin{subfigure}{0.45\textwidth}
\includegraphics[width=1\linewidth, height=4cm]{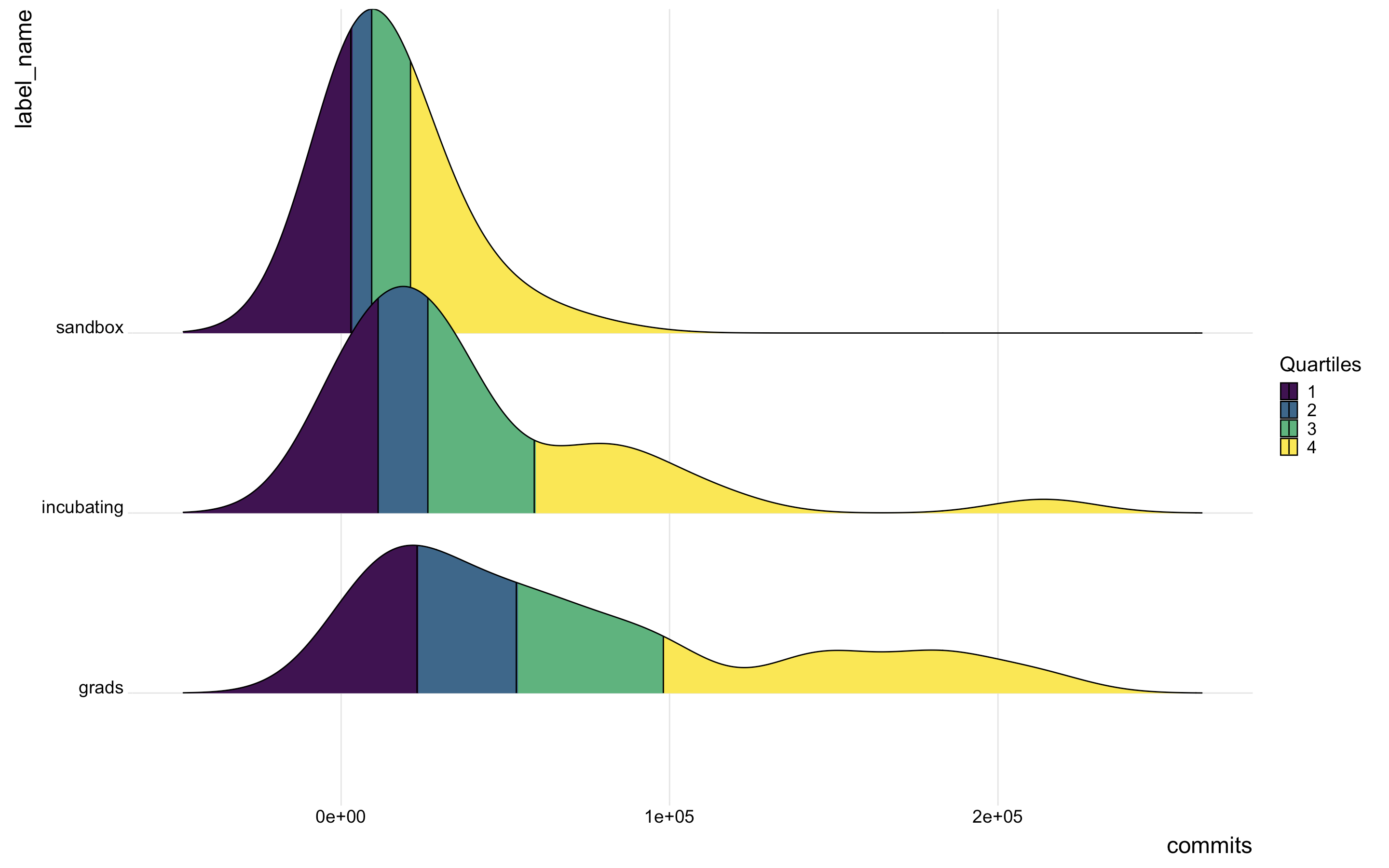}
\caption{Commits}
\label{fig:subim4}
\end{subfigure}

\begin{subfigure}{0.3\textwidth}
\includegraphics[width=1\linewidth, height=4cm]{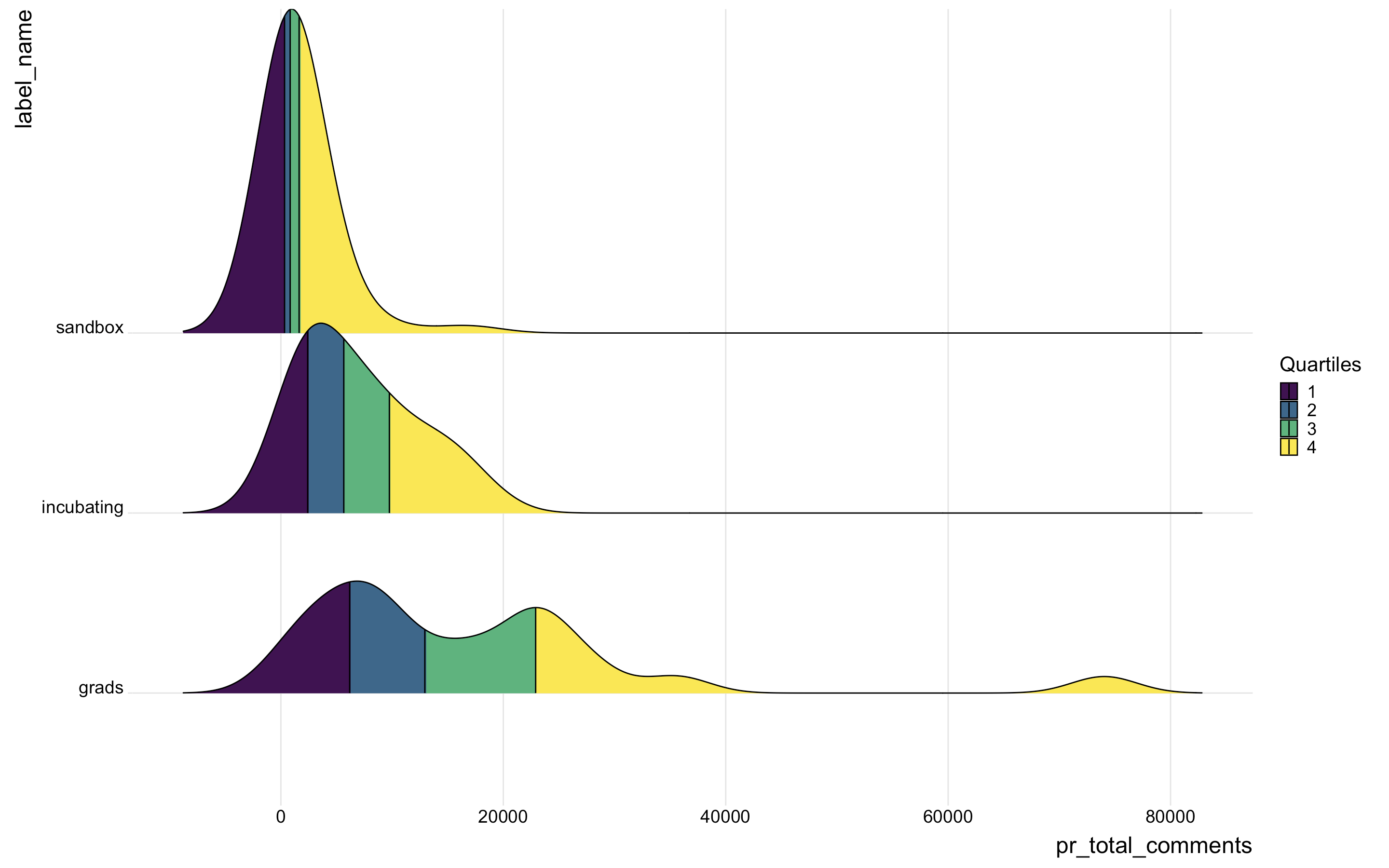}
\caption{Pull request total comments}
\label{fig:subim5}
\end{subfigure}
\begin{subfigure}{0.3\textwidth}
\includegraphics[width=1\linewidth, height=4cm]{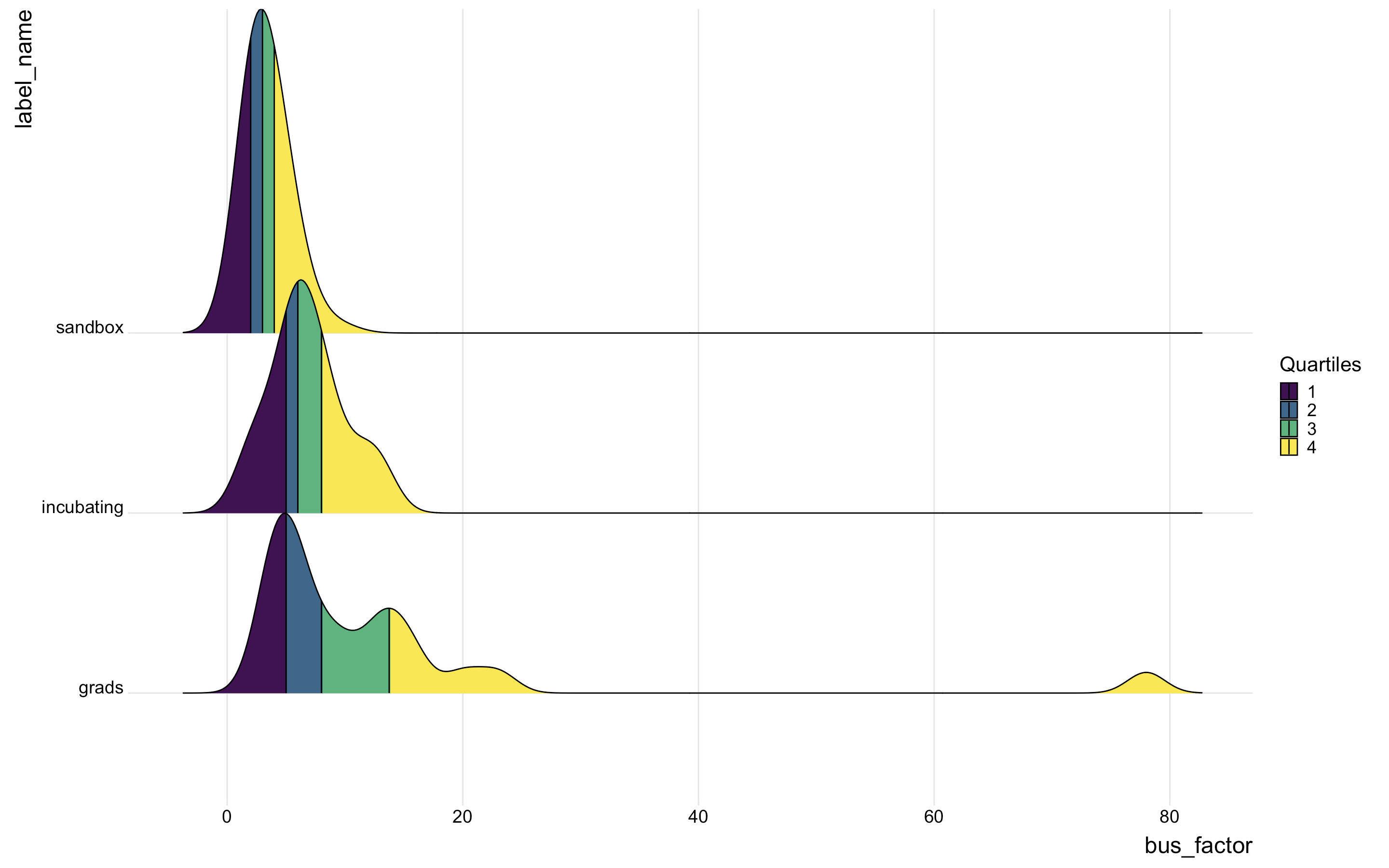}
\caption{Bus factor}
\label{fig:subim3}
\end{subfigure}
\begin{subfigure}{0.3\textwidth}
\includegraphics[width=1\linewidth, height=4cm]{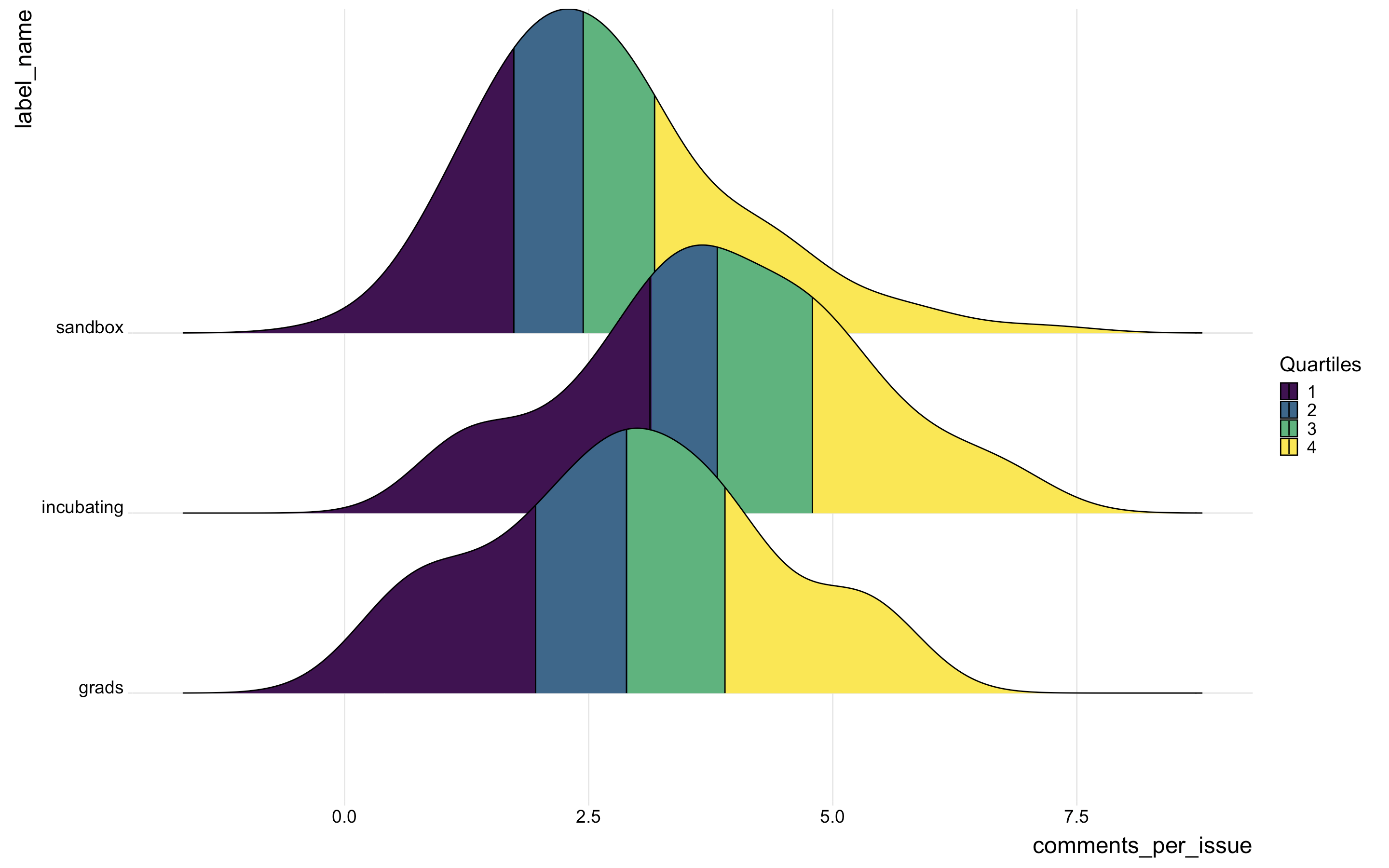}
\caption{Comments per issue}
\label{fig:subim4}
\end{subfigure}

\caption{Ridge line plots for the seven factors in the model for classifying projects according to their lifecycle stage that were not discussed in the core narrative of the paper.}
\label{fig:image2}
\end{figure}

\end{document}